%% file: main.tex
\documentclass[10pt,pre,twocolumn,tightenlines,longbibliography,notitlepage, nofootinbib,superscriptaddress]{revtex4-2}
\pdfoutput=1

\usepackage{amsmath, amssymb}
\usepackage{xcolor}
\usepackage{verbatim}
\usepackage{graphicx}
\usepackage{natbib}
\usepackage{siunitx}
\usepackage{upgreek}
\usepackage[textsize=tiny]{todonotes}
\usepackage{longtable}

\definecolor{light-gray}{gray}{0.95}

\begin{document}

\title{Measuring multisubunit mechanics of geometrically-programmed colloidal assemblies via cryo-EM multi-body refinement}

\author{Thomas E. Videb\ae k}
\email{videbaek@brandeis.edu}
\affiliation{Martin A. Fisher School of Physics, Brandeis University, Waltham, MA, 02453, USA}
\author{Daichi Hayakawa}
\affiliation{Martin A. Fisher School of Physics, Brandeis University, Waltham, MA, 02453, USA}
\author{Michael F. Hagan}
\affiliation{Martin A. Fisher School of Physics, Brandeis University, Waltham, MA, 02453, USA}
\author{Gregory M. Grason}
\affiliation{Department of Polymer Science and Engineering, University of Massachusetts, Amherst, Massachusetts 01003, USA}
\author{Seth Fraden}
\affiliation{Martin A. Fisher School of Physics, Brandeis University, Waltham, MA, 02453, USA}
\author{W. Benjamin Rogers}
\email{wrogers@brandeis.edu}
\affiliation{Martin A. Fisher School of Physics, Brandeis University, Waltham, MA, 02453, USA}

\begin{abstract}
 Programmable self-assembly has recently enabled the creation of complex structures through precise control of the interparticle interactions and the particle geometries. Targeting ever more structurally complex, dynamic, and functional assemblies necessitates going beyond the design of the structure itself, to the measurement and control of the local flexibility of the inter-subunit connections and its impact on the collective mechanics of the entire assembly. In this study, we demonstrate a method to infer the mechanical properties of multisubunit assemblies using cryogenic electron microscopy (cryo-EM) and RELION's multi-body refinement. Specifically, we analyze the fluctuations of pairs of DNA-origami subunits that self-assemble into tubules. By measuring the fluctuations of dimers using cryo-EM, we extract mechanical properties such as the bending modulus and interparticle spring constant. These properties are then applied to elastic models to predict assembly outcomes, which align well with experimental observations. This approach not only provides a deeper understanding of nanoparticle mechanics, but also opens new pathways to refining subunit designs to achieve precise assembly behavior. This methodology could have broader applications in the study of nanomaterials, including protein assemblies, where understanding the interplay of mechanical properties and subunit geometry is essential for controlling complex self-assembled structures.
\end{abstract}

\maketitle

In recent years, the field of programmable self-assembly has demonstrated the ability to create increasingly complex structures from increasingly complex building blocks, ranging from DNA-based colloidal particles~\cite{liu2016diamond, tian2016lattice,tikhomirov2018triangular, he2020colloidal,jacobs2024assembly} to de novo-designed proteins~\cite{de2017antimicrobial, hsia2021design, li2023accurate, lutz2023top}. These advances are enabled by newfound control over the specificity of the interactions in multicomponent mixtures~\cite{zeravcic2014size, jacobs2015rational, rogers2020mean, romano2020designing, hayakawa2024symmetry, liu2024inverse} and the relative orientations of bound subunits, which are encoded through a combination of the subunit geometries and valence-limited, directional interactions~\cite{sigl2021,Hayakawa2022,posnjak2024diamond}. The prevailing paradigm in these programmable assemblies is to define geometrically rigid particles and interactions compatible with a target structure.  Going beyond this, towards engineering functional assemblies that mimic the remarkable properties achieved, for example, by multi-protein complexes in biology, requires control over not just the target structure, but also the {\it deformability} of assemblies on the multi-unit scale and how these are controlled by the deformability of interactions at the inter-unit scale.  Intuitively, the fidelity of targeting specific geometrically programmed structures is inevitably limited by the compliance of inter-particle interactions at the pairwise scale.  Additionally, the effects of intra-assembly stresses have been proposed as a means to rationally harness geometrical frustration to achieve self-limiting architectures~\cite{grason2016perspective}, and allostery-like, stimuli-responsive propagation of cooperative intra-assembly motions remains a critical engineering challenge for programmable assembly~\cite{gerling2015dynamic,marras2015programmable,pumm2022dna}. However, at present, quantitative approaches that characterize and are predictive of the deformability of inter-subunit geometry, and its ultimate impacts on the collective mechanics of multisubunit assemblies, are lacking.

Recent advances in cryogenic electron microscopy (cryo-EM) provide a path forward to characterizing the dynamic processes and conformational flexibility of nanoscopic particles and their assemblies. These developments are driven by new computational tools, such as RELION's multi-body refinement~\cite{nakane2018characterisation}, DynaMight~\cite{schwab2024dynamight}, 3DFlex~\cite{punjani20213d}, and cryo-DRGN~\cite{kinman2023uncovering}, which enable the characterization of sample heterogeneity,  both for large movements of relatively rigid regions and for continuous deformations. Along with time- and temperature-resolved cryo-EM methods, these approaches have so far been used to learn about various conformational states~\cite{xu2024conformational,li2024cryo} and dynamic and structural changes related to the functions of enzymes~\cite{torino2023time,chen2019temperature,hu2024physiological}. While there have been some studies examining distributions of distinct states in small complexes~\cite{harris2023energetic,zhang2024angle}, the development of general, cryo-EM-based frameworks to characterize dynamical conformations to quantify and design the mechanics of subunits and their assemblies is still in its infancy.

\begin{figure*}[!th]
    \centering
    \includegraphics[width=\linewidth]{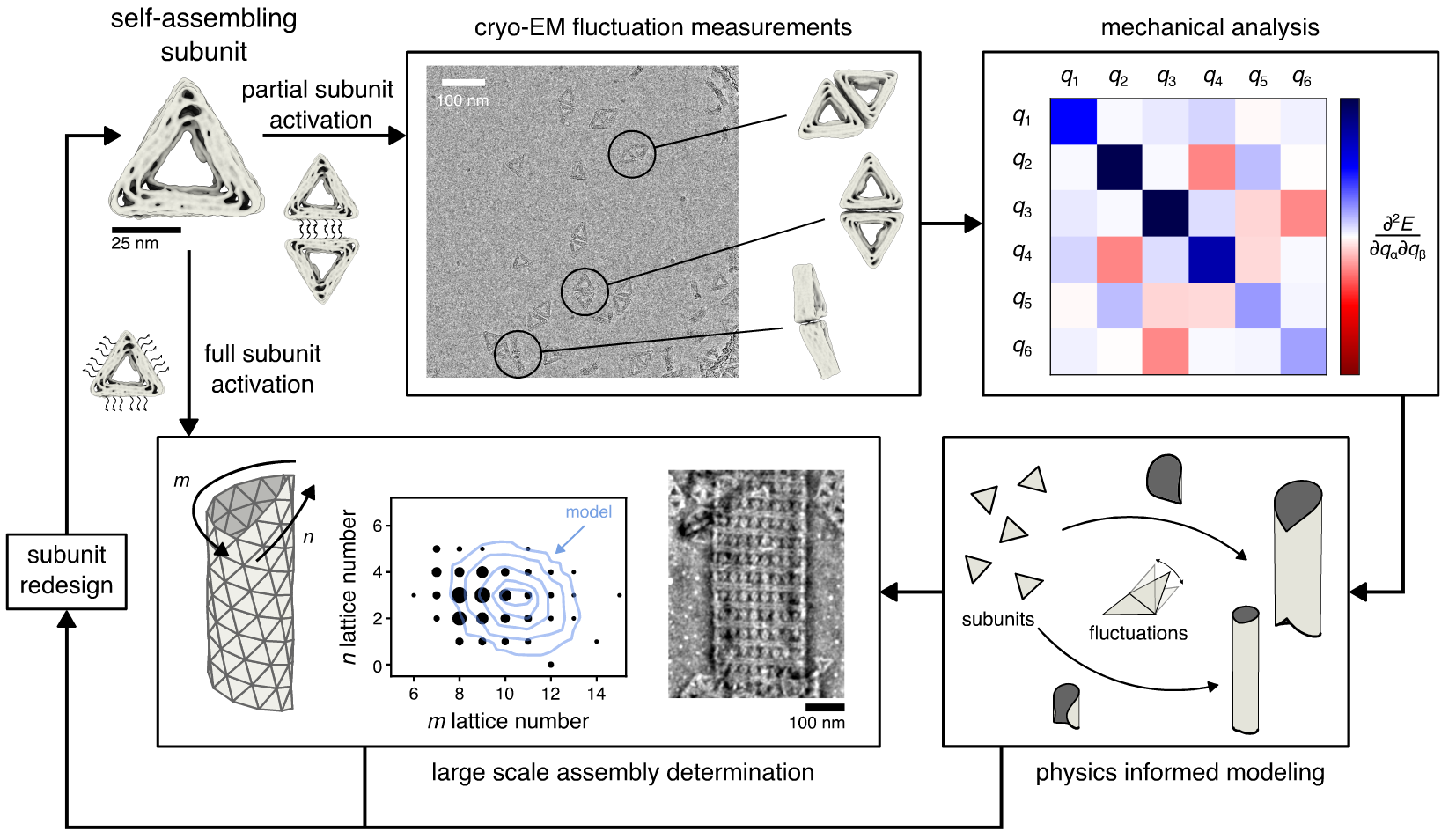}
    \caption{\textbf{Inferring mechanical properties of self-assemblies using cryo-EM}. Using a DNA-origami monomer we can fully activate or partially activate the subunit. With full activation, the subunit will assemble into a distribution of self-limited tubules. With partial activation, dimers will form. Using cryo-EM, we can characterize the dimer’s fluctuations and extract mechanical properties such as the bending modulus or interparticle spring constant. These mechanical properties are input into an elastic energy model that can then predict the assembly of the monomer upon full activation. This information can then be used to iteratively redesign the monomer to achieve a desired goal.}
    \label{Fig:fig-overview}
\end{figure*}

In this work, we utilize cryo-EM to characterize the individual particles, the intersubunit mechanics, and small assemblies of nanoscale colloidal particles, and show how this information can be used to understand and rationally control their cooperative assembly. We fabricate and study protein-like DNA-origami objects that self-assemble into tubules via user-specified control over their local curvature and specific interactions. This type of self-limited assembly is susceptible to polymorphism, e.g., tubules of varying circumference and helicity, that results from thermally-excited bending fluctuations in the growing assembly~\cite{fang2022polymorphic,Hayakawa2022,videbaek2024economical}. For tubule-forming monomers, we measure the fluctuations of dimers using cryo-EM and multi-body refinement, and extract the associated mechanical properties. These mechanical properties are then used as inputs to a series of elastic models of tubule assembly of increasing complexity. Crucially, we find that overly simplistic models that consider only the single-subunit geometry or the pairwise bending mechanics fail to describe the experimentally observed distributions of tubule assembly. In contrast, a model that includes the cooperative motions of multisubunit structures matches our experiments quantitatively, highlighting the importance of being able to characterize and control the collective mechanics. We conclude by showing how insights from our approach can be used to rationally redesign the monomers to achieve a desired assembly outcome (Fig.~\ref{Fig:fig-overview}).

\section*{Results}

\begin{figure*}[!th]
    \centering
    \includegraphics[width=\linewidth]{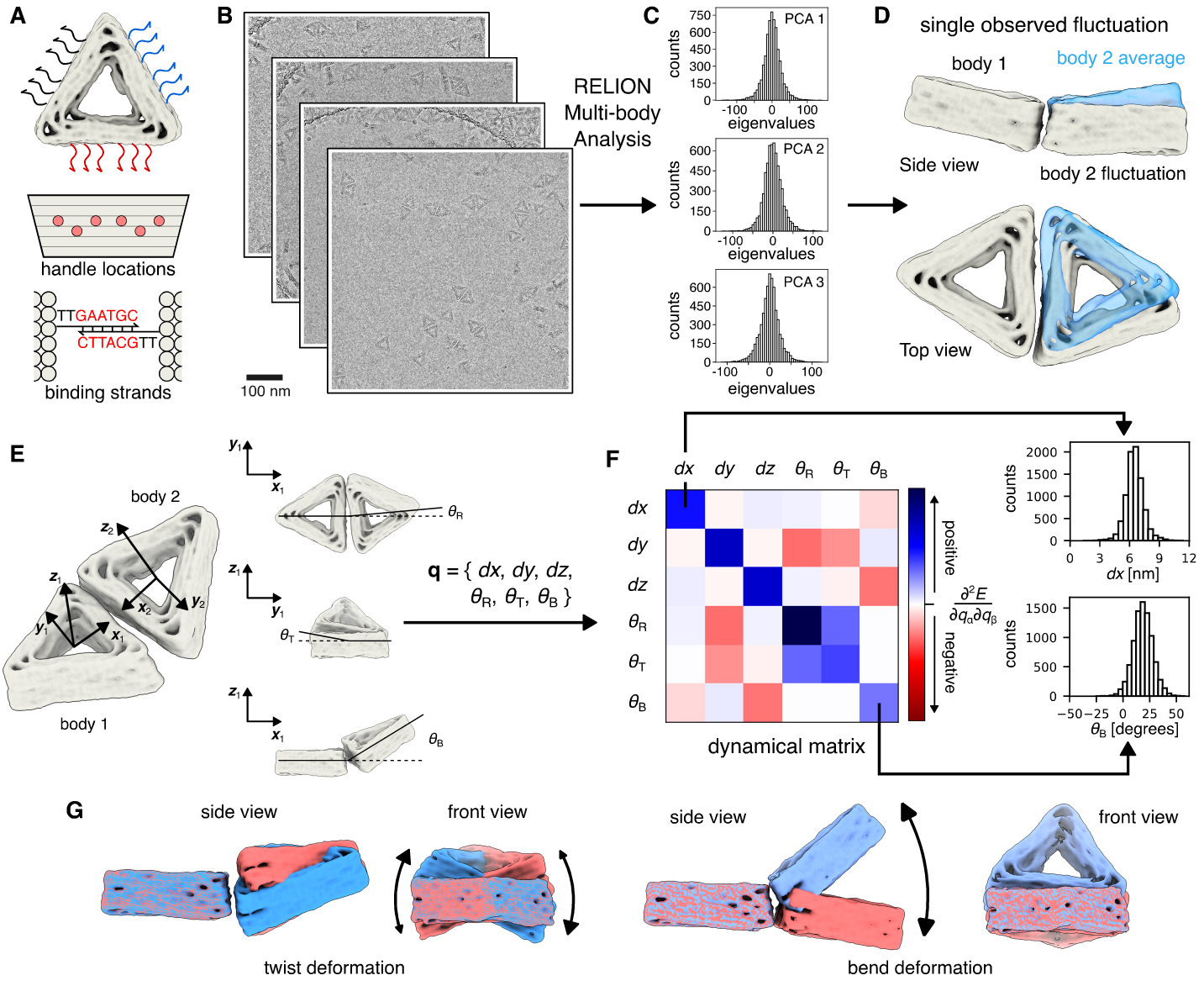}
    \caption{ \textbf{Mechanical moduli extraction using cryo-EM.} (A) DNA-origami monomer with beveled edges and single-stranded DNA extruded from the edges to mediate interactions. Each side has six binding strands coming out of the third and fourth helix. (B) Cryo-EM images of dimers are used to get a reconstruction of the average dimer configuration. (C) RELION multi-body refinement is used to calculate the principal components of motion for the dimer. (D) Individual fluctuations observed in the cryo-EM images can then be generated from the multi-body output. (E) For each observed fluctuation we define a set of coordinates for each body with respect to their mutual interaction site. From these, we extract relative translations ($dx,\ dy,\ dz$) and angles of the two bodies, what we call roll, $\theta_\mathrm{R}$, twist, $\theta_\mathrm{T}$, and bend, $\theta_\mathrm{B}$. (F) Relative body coordinates are used to construct a dynamical matrix that contains mechanical information about the dimer. (G) Representative particle renderings that show the extremes of twist and bend deformations observed.}
    \label{Fig:fig-method}
\end{figure*}

We use an experimental system consisting of DNA-origami triangular subunits that have a user-programmable geometry and interaction specificity, which we exploit to make large-scale self-closing assemblies, as well as small sub-assemblies. This class of synthetic subunits, developed to share many attributes with the protein building blocks of life, such as their complex geometry and valence-limited, specific interactions, has been used to make analogs of viral capsids~\cite{sigl2021, wei2024} and microtubules~\cite{Hayakawa2022,videbaek2024economical}. For this study, we use a single monomer design with three distinct edges that have been beveled to target the self-assembly of cylindrical tubules. Each edge is decorated with single-stranded DNA (ssDNA) `sticky ends' that mediate specific subunit-subunit interactions (Fig.~\ref{Fig:fig-method}A). In particular, six binding strands extend from about the mid-plane of the sides of the subunits, each of which is composed of a unique six-base-pair binding domain and a two-nucleotide spacer domain (Fig.~\ref{Fig:fig-method}A) (see Table S1 for sequence information). Figure~\ref{Fig:fig-method}A shows the cryo-EM single-particle reconstruction of the monomer. This monomer is the same as the one used in a previous study~\cite{videbaek2024economical}.

To study the mechanics of our subunits, we characterize the conformational distribution of pairs of bound particles, which we call `dimers', using cryo-EM and multi-body refinement. While one could look at the fluctuations within a single subunit~\cite{ni2022direct}, doing so would ignore any effects that could arise from the flexibility of the interparticle contacts. This consideration is especially relevant for our experimental system because the subunits themselves are effectively rigid while the rigidity of the interparticle connections is comparatively more flexible. Recent work on DNA-origami bundles shows that tight packings of helices similar to the body of our triangle design can have persistence lengths up to several micrometers~\cite{castro2015mechanical,lee2019tailoring}. In contrast, single- and double-stranded DNA only have persistence lengths of $\sim$2 nm and $\sim$50 nm respectively~\cite{murphy2004probing,geggier2011temperature}. This disparity suggests that the single-stranded portion of the DNA handles is the most compliant element in bound pairs, and therefore, is likely determinative of the mechanical response of assemblies.  

\subsection*{Characterizing elastic properties with cryo-EM}

To sample the space of fluctuations accessible to our interacting subunits, we collect a set of dimer images with cryo-EM and analyze them using RELION’s multi-body analysis. We create dimers by activating only a single edge of our monomer with ssDNA interactions: The handles are removed from the other two edges. These dimers are then allowed to equilibrate at a constant temperature in buffer (20 mM MgCl$_2$, 1xFoB, see Methods) before vitrification and subsequent imaging. From a collection of particle orientations (Fig.~\ref{Fig:fig-method}B), we reconstruct the average conformation of the dimer. Since we are interested in quantifying the fluctuations about this average structure, we use multi-body refinement~\cite{nakane2018characterisation}, assuming that each monomer is a rigid body, to get the principle components of the fluctuations (Fig.~\ref{Fig:fig-method}C). To understand the implications of the most complaint modes of inter-particle fluctuation for assembly mechanics, it is necessary to relate principle coordinates to a local basis of inter-particle spatial and orientational distortions.

To gain mechanistic insight into dynamic structures using cryo-EM, we perform a mechanical analysis of the fluctuations using coordinates defined relative to the position and orientation of triangular particles~\cite{hicks2010coarse}. From the principle components, we can take any observed fluctuation (Fig.~\ref{Fig:fig-method}D)(see SI Appendix, Section 1 for details) and cast it into a local-coordinate system defined by the pair of triangles (Fig.~\ref{Fig:fig-method}E). In this study, we use triangular DNA-origami subunits that can interact through edge-to-edge binding. Therefore, it is natural to choose a coordinate system for each body that is oriented with respect to the interacting edge. With these coordinate systems defined, we measure relative displacements and rotations of the bodies, what we call roll, twist, and bending angles (SI Appendix, Section 2). This approach gives us a final set of differential coordinates for each observed conformation of the dimer. We note that analogs of this type of edge-edge coordinate system may be useful in the future to analyze the fluctuations of proteins that interact through beta-sheets.

After measuring the differential coordinates for all of the fluctuations that we observe, we infer the mechanical properties of the dimer by calculating the dynamical matrix from the covariance of the differential coordinates. The dynamical matrix is defined as 
\begin{equation}
M_{q_\alpha q_\beta} = k_{\rm B}T \mathrm{cov}(q_\alpha, q_\beta)^{-1}
\end{equation}
where $q_\alpha$ is a component of the 6-dimensional deformation vector ${\bf q}$ of a fluctuation (Fig.~\ref{Fig:fig-method}F). By looking at projections of the fluctuations onto any particular coordinate we find that the fluctuations are normally distributed (Fig.~\ref{Fig:fig-method}F). Due to the Gaussian nature of these fluctuations, we can assume that they result from thermal fluctuations and are characterized by a Boltzmann distribution from the following elastic energy, 
\begin{equation}
\label{eq: deltaF}
\Delta E ({\textbf{q}}) = \frac{1}{2} (q_\alpha - q_{\alpha,0}) M_{q_\alpha q_\beta}(q_\beta - q_{\beta,0})
\end{equation}
which is related to the probability distribution for a given deformation by $P({\bf q}) \propto  \exp \big[ -\Delta E ({\bf q})/k_{\rm B} T \big]$.  Because $M$ is not a diagonal matrix (Fig. ~\ref{Fig:fig-method}F), the components of ${\bf q}$ (e.g. bend, twist, and roll) are not normal modes.  It is nevertheless, useful to compare the relative magnitudes of the distinct diagonal elements, including the effective angular range of each type of motion, as shown in Fig.~\ref{Fig:fig-method}G. The off-diagonal elements of $M$ reveal cross-correlations between motions.

\begin{figure*}[!th]
    \centering
    \includegraphics[width=\linewidth]{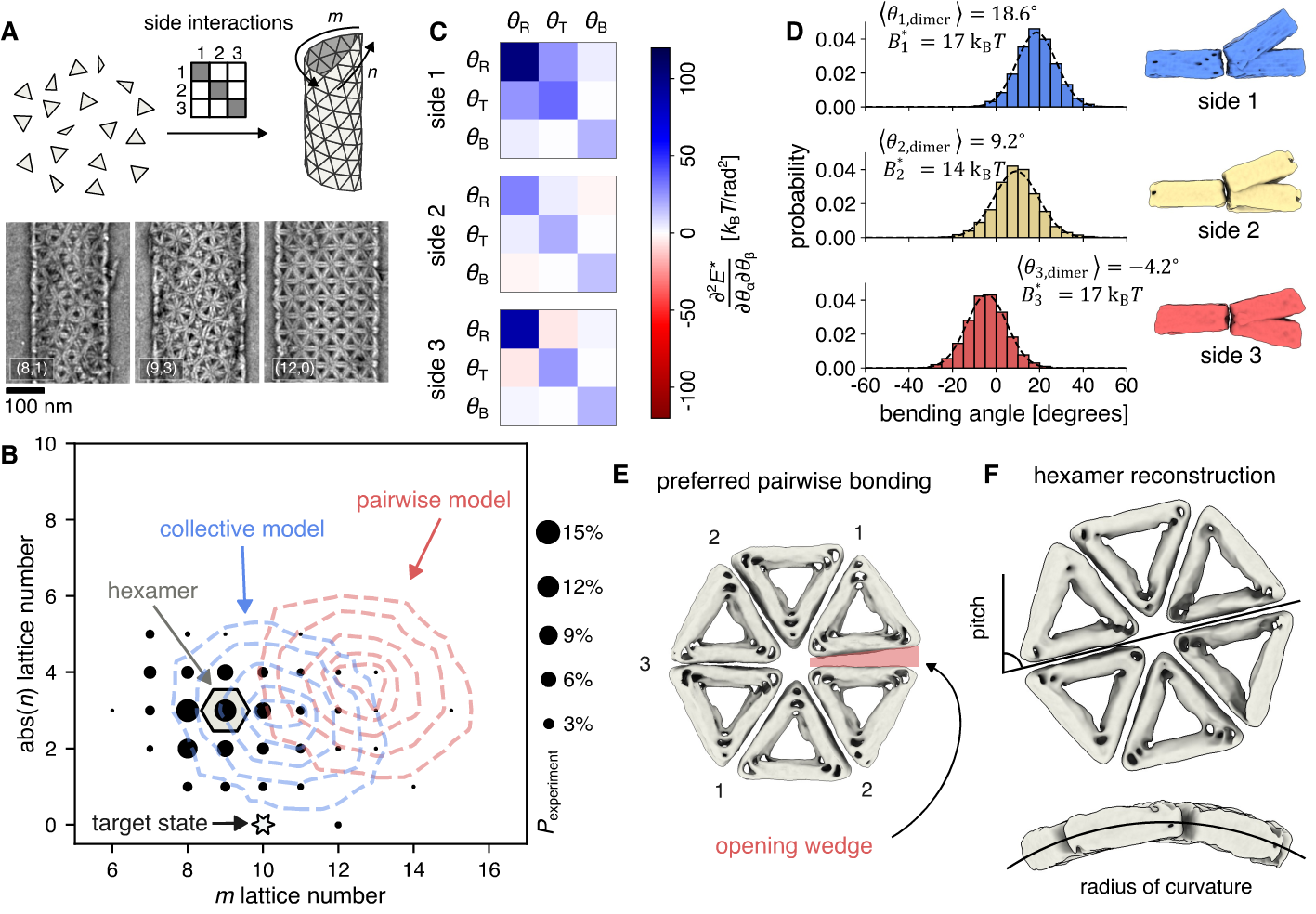}
    \caption{\textbf{Fluctuations of tubule dimers.} (A) Schematic of tubule assembly using triangular subunits. Representative TEM micrographs of assembled tubules are shown. Scale bar, 100 nm. Micrographs have been bandpass filtered to improve contrast. (B) Tubule state distribution where each point represents a single (m,n) tubule state and the size shows the relative probability of observing that state, experimental data is taken from ref.~\cite{videbaek2024economical}. Note, in TEM we can only measure the absolute value of $n$. Dashed lines show iso-contours of probability from our elastic energy models at levels of 10\%, 25\%, 50\%, 75\%, and 90\% of the maximum distribution probability; $P_\mathrm{max,pairwise}=11.1\%, P_\mathrm{max,collective}=11.5\%$. The shaded hexagon shows the state corresponding to the most likely tubule stat inferred from an average cryo-EM hexamer configuration. (C) Dynamical matrices for the measured angle fluctuations for each of the dimers. (D) Distributions of the opening angles for dimers of each side of the tubule monomer were measured with cryo-EM. Each distribution is fit to a Gaussian from which we extract the average opening angle and an effective bending modulus. Images on the right show the extent of bending fluctuations. (E) An example of the incompatibility of measured pair-wise interactions on forming a hexamer loop. Based on the dimer measurements, we placed monomers and their neighbors for all but the last bond in a hexamer loop. Numbers denote which side interaction is at each binding site. At the last binding edge, there is a gap. (F) Cryo-EM reconstruction of a hexamer of monomers. We fit the electron density map to a cylindrical surface and extract both the radius of curvature and pitch of the hexamer (SI Appendix, Section 5).}
    \label{Fig:fig-tubules}
\end{figure*}

Before continuing, we have to account for some aspects of the cooling of the sample during vitrification. Prior to liquid ethane plunging, we prepare samples of dimers that equilibrate at room temperature, but a question remains as to whether or not the same equilibrium distribution remains throughout cooling. Typical cooling rates in liquid ethane are around $10^{6}-10^{8}$ K/s~\cite{dubochet1988cryo}, meaning it takes about $1-100~\upmu$s for a sample to go from room temperature to the vitrification temperature of water ($\sim$130 K). For nanoscale objects (our particles are $\sim$50 nm in size) the rotational-diffusion timescale is on the order of nanoseconds. This means that during the cooling process, the distribution of states will in all likelihood be able to re-equilibrate, resulting in a narrower distribution of states than one would expect at room temperature. This type of behavior has been reported in molecular dynamics simulations of proteins~\cite{bock2022effects}. As an initial correction to account for this cooling process, we rescale the measured moduli from the vitrification temperature up to room temperature, $M^* = M(T_\mathrm{vitrification}/T_\mathrm{room})$. We return to this assumption in the Discussion section.

As an example of how to interpret the fluctuation data, we look at the dynamical matrix shown in Fig.~\ref{Fig:fig-method}F. First, we look at the distributions of the coordinates: See the histograms in Fig.~\ref{Fig:fig-method}F for the distributions of the displacement $dx$ and the bending angle $\theta_\mathrm{B}$. These two coordinates show the smallest degree of cross-coupling with other modes and both are well-approximated by a Gaussian distribution. Using the intuition we built from Eqn.~2, we compute the effective moduli of these modes, finding $k_x^* = 0.53\ k_\mathrm{B}T/\mathrm{nm}^2$ and $B^*=17.5\ k_\mathrm{B}T/\mathrm{rad}^2$. While the stretching modulus is smaller than expected, the bending modulus is in good agreement with an entropic model for DNA linkers, providing a convenient sanity check for our analysis framework (SI Appendix, Section 3). Next, we look at the structure of the dynamical matrix. Focusing on the bottom corner we note two features: 1) the bend angle is independent of twist and roll; and 2) twist and roll are coupled together. While these observations are interesting in their own right, connecting them to the downstream outcomes of self-assembly---an essential step in the rational design of programmable self-assembling materials---further requires a modeling framework for relating the mechanics to the distribution of structures one might expect to form in an experiment.

\subsection*{Fluctuations in tubule assemblies}

In the self-assembly of tubules from a single species of triangular monomer, each side of the monomer needs to have a well-specified binding angle to precisely target a tubule with a particular width and helicity. By encoding self-homologous interactions on each side of the monomer, the encoded binding angles result in an accumulation of curvature during assembly, leading to self-closure (Fig.~\ref{Fig:fig-tubules}A). Here, we study the DNA-origami monomer from ref.~\cite{videbaek2024economical} that was designed to assemble into an achiral tubule with 10 monomers in circumference. However, fluctuations in the binding angles allow for a distribution of tubule states with varying diameters and helicities. We characterize this distribution by measuring the propensity of forming any given tubule state (Fig.~\ref{Fig:fig-tubules}B). A single state can be denoted by a pair of indices, ($m,n$), that describes the shortest self-closing loop that can be made while following the edges of the triangular lattice. There are two aspects of this distribution we hope to uncover: 1) What is the source of the breadth of the distribution?; and 2) Why does the center of the distribution deviate from the targeted (10,0) state? Potentially the latter aspect could be addressed by single-particle cryo-EM, but this involves assumptions of how other particles will come to bind at the interface and has led to inaccurate predictions in the past~\cite{Hayakawa2022}.

We now aim to understand whether or not our dimer-level characterization can account for the polymorphism we observe in our experiments, and therefore provide a detailed understanding of its origin. Since each side of the monomer has a unique binding angle in general, we characterize the fluctuations of each side by making three different dimer samples. After taking cryo-EM data for dimers of each side and performing multi-body refinement, we compute the dynamical matrix between the angular coordinates for the dimers (Fig.~\ref{Fig:fig-tubules}C). Looking at the diagonal elements of these matrices, we see that the element corresponding to the bending angle, $\theta_\mathrm{B}$, has the smallest value as well as the lowest degree of coupling to the other angles. This observation gives an inclination that perhaps a simple elastic-energy model that only accounts for pair-wise bending fluctuations~\cite{Helfrich1988, videbaek2022tiling, Hayakawa2022} could account for the distribution of states that we observe.

To begin, we extract the bending angles for each dimer (Fig.~\ref{Fig:fig-tubules}D). We see that each side has an angle distribution that can be well-described by a Gaussian, hinting that the fluctuations are due to purely thermal motion. From these distributions, we can measure the average binding angle for each side, as well as estimate their bending moduli from the variance of each distribution: $B = \mathrm{var}(\theta_{\rm B})^{-1}$.

Taking the measured angles and bending rigidities as inputs, we first show that a simple elastic model with only pairwise bending fluctuations, decoupled from the other deformation modes, fails to capture the experimentally measured distribution of tubule geometries. Assuming the main contribution to the probability of forming a tubule comes from the elastic bending energy of the membrane (assuming stretching is negligible)~\cite{fang2022polymorphic}, we compute the energy of each unique tubule state by evaluating
\begin{equation}
    E_\mathrm{pairwise}=\frac{1}{4} N(m,n) \sum_{i\in\mathrm{sides}} B_i \big(\theta^{(i)}_{\rm B}(m,n) - \theta^{(i)}_{{\rm B},0} \big)^2,
\end{equation}
where $N(m,n)$ is the number of components in the assembly, $B_i= M^{(i)}_{\theta_{\rm B} \theta_{\rm B}}$ is the bending rigidity of side $i$, $\theta^{(i)}_{\rm B}(m,n)$ is the bending angle of side $i$ in an ($m,n$) tubule, and $\theta^{(i)}_{{\rm B},0}$ is the preferred bending angle of side $i$. For $N(m,n)$ we take the number of monomers at the point of closure~\cite{fang2022polymorphic} because, once a tubule begins to extend, the probability of opening a tubule into a sheet to change its type goes to roughly zero. We then assume that the tubule states follow a Boltzmann distribution based on this elastic energy (see SI Appendix, Section 4 for more details). Figure~\ref{Fig:fig-tubules}B shows the predictions of this model using the measured $B$ and $\theta_{{\rm B},0}$ from cryo-EM as red contours overlaid on the experimental measurements. It is clear from this comparison that this simple model is insufficient to explain the origin of the experimental observations: Although it roughly captures the spread, it fails to capture the center of the distribution entirely.

We next show that the discrepancy between the decoupled pairwise picture and the experiment arises from a combination of the interrelation of pair-wise distortions in the triangular lattice and the elastic cross-couplings between angular modes at each edge.  We first illustrate that it is not possible to satisfy the locally preferred pairwise inter-subunit displacements in a hexagonal sheet, by constructing a single hexameric loop, shown in Fig.~\ref{Fig:fig-tubules}E, in which each of five of the six interparticle position- and angle-displacements favored by each edge are perfectly satisfied (i.e. ${\bf q} = {\bf q}_{0}$ for all edges).  Notably, around this closed hexamer loop, these displacements are {\it geometrically incompatible}, as evidenced by the opened wedge on the right of the hexamer. (Fig.~\ref{Fig:fig-tubules}E). With the additional degrees of freedom added by twist and roll angles, we can no longer define a single geometry for a given ($m,n$) state, and hence cannot compute the elastic energy.

To make progress, we turn to simulation to find the minimum energy configuration of triangles in an ($m,n$) tubule given the preferred average displacements and binding angles measured with cryo-EM. We initialize triangles in an ($m,n$) tubule configuration using our bending-angle model~\cite{Hayakawa2022} and make trial displacements and rotations on the triangles until a new configuration is reached that minimizes the total elastic energy of all pairwise interactions within the structure
\begin{equation}
    E_{\rm collective}= \frac{1}{2} N(m,n) \sum_{i\in \rm sides} \Delta E^{(i)}(\bf q),
\end{equation}
where $\Delta E^{(i)}(\bf q)$ is elastic energy for side $i$ derived from the full dynamical matrix we measured, Eqn.~\ref{eq: deltaF}. From the energy-minimized configuration, we can then calculate the average bulk elastic energy per monomer in a given tubule state (see SI Appendix, Section 4B for details). Using these monomer energies, we then assume, again, that the tubule states follow a Boltzmann distribution and compute a model distribution based on these compatible tubule geometries, shown in Fig.~\ref{Fig:fig-tubules}B as the blue contours.

The elastic energy model with compatible tubules, which we call the ``collective model'', shows strong quantitative agreement with experimental observations. While the center of the calculated distribution does not exactly match the experimental distribution, it is much closer than our initial pairwise model. Also, the breadth of the distribution matches well, both showing a five-fold drop in the probability of states that are two to three lattice steps away from the center of the distributions.

An implication of the shift in the mean distribution when including twist and roll is that the average binding angles within an assembled tubule are different from the binding angles of the dimers alone. In other words, multiple subunits need to cooperatively interact, reorient, and relax. To explicitly observe this multisubunit cooperativity, we use cryo-EM to reconstruct a closed hexamer of subunits (Fig.~\ref{Fig:fig-tubules}F), which is made stable by only activating the subunit binding interactions on the intra-hexamer subunit edges. We chose this geometry because it is the smallest assemblage that can be expected to mimic the subunit orientational constraints arising from the lattice in a tubule. We find that indeed the hexamer forms a patch of a cylindrical shell: The hexamer is flat in one direction and curved in the orthogonal direction. We then fit the electron density map to a cylindrical surface to extract both the radius of curvature as well as the pitch of the triangles within the patch (SI Appendix, Section 5). This fit gives a radius of curvature of 1.81$l_\mathrm{mono}$ and a pitch of 3.19$l_\mathrm{mono}$, where $l_\mathrm{mono}$ is the lattice spacing of monomers in the tubule and was measured to be 68.7~nm. These hexamer measurements correspond to a closest tubule state of (9,3). This inferred tubule state is in excellent agreement with the most likely assembled tubule states from our experiment.

\begin{figure}[!t]
    \centering
    \includegraphics[width=\linewidth]{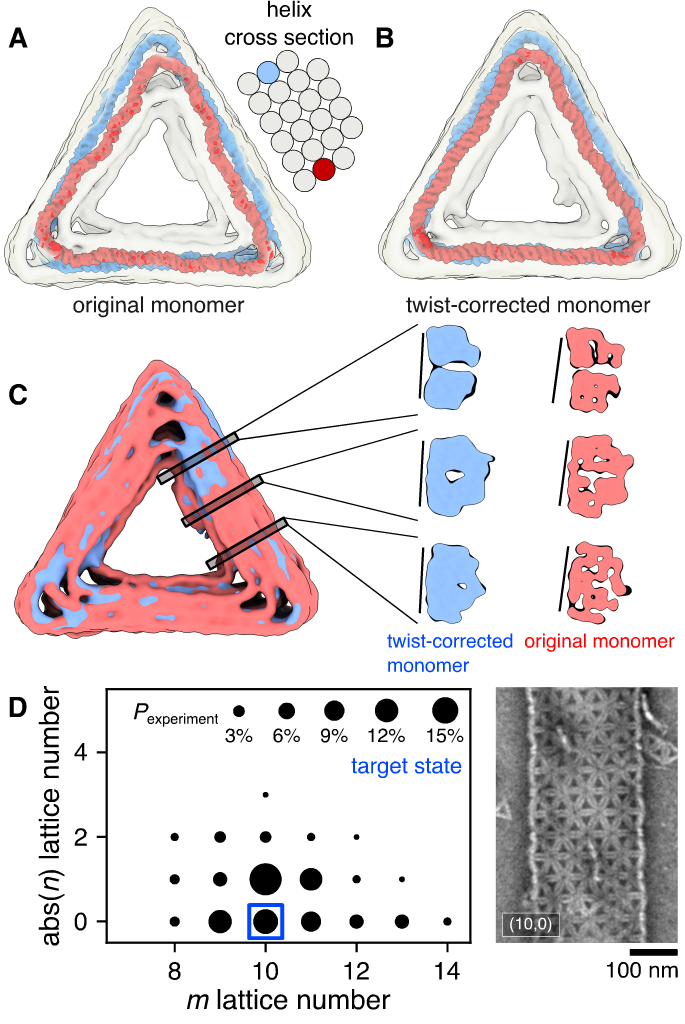}
    \caption{\textbf{Redesigning subunit geometry.} (A) Cryo-EM map of the initial monomer with two helices from the inner and outer layers of the monomer colored in, the cross-section of the helices denotes which ones are shown. (B) Cryo-EM map of the twist-corrected monomer with inner and outer helices colored in. (C) Overlapping cryo-EM maps of the original subunit (red) and the redesigned subunit to remove the excess twist (blue). Below, cross-sections along an edge are shown with lines of constant angle as guides for the eye. The original monomer shows considerable twist along its edge while the redesigned monomer maintains a consistent angle. (d) Tubule distribution for the twist corrected monomer. Since we have to plot abs($n$), states corresponding to $n\neq0$ may contain both positive and negative $n$ states.}    
    \label{Fig:fig-redesign}
\end{figure}

\subsection*{Iterative feedback for monomer design}
Now that we have a more complete picture of how subunit geometry and mechanics combine to control assembly, we can refine the initial design of the subunit. To more accurately assemble the desired (10,0) tubule, for example, we need to both remove the non-zero twist and roll angles from the monomer and achieve the correct angles on each side. In looking more closely at the single-particle reconstruction of our original monomer, we find that the top and bottom layers of the triangle are twisted with respect to each other (Fig.~\ref{Fig:fig-redesign}A), and upon inspection of the initial DNA-origami design, we observe that we failed to include a high enough skip density to correct for the overwinding of DNA on the square lattice of the DNA origami~\cite{ke2009multilayer}. 

Notably, adjusting the monomer to eliminate the twist yields a subunit that assembles into the desired target structure. First, we look at the single particle cryo-EM reconstruction to check that the twist along the sides has been corrected. The new design shows a reduced twist of the top and bottom layers when compared to the initial monomer (Fig.~\ref{Fig:fig-redesign}B). When we overlap the electron density maps and look at a cross-section of one of the sides (Fig.~\ref{Fig:fig-redesign}C), we see that while the old design noticeably changes the angle along the length of one side, the new monomer has a consistent angle. Next, we look to see if the redesign changed the distribution of assembled tubules in a meaningful way. Figure~\ref{Fig:fig-redesign}D shows the distribution for the twist-corrected monomer, which is now centered around (10,0), as intended. We note that even though the (10,1) state looks like it is the most populated state, this data point contains both tubules with $n=\pm 1$ because transmission electron microscopy cannot differentiate between right- and left-handed tubules. The success of this data- and model-driven redesign of our triangular monomer is a testament to the potential of the characterization and analysis framework we developed above.

\section*{Discussion}

In this report, we showed how multi-body analysis of cryo-EM data can be used to quantitatively extract the mechanical properties of the intersubunit connections and small assemblies of nanometer-scale colloidal particles. We developed an analysis procedure that expands the qualitative information of principal component analysis to the quantitative aspects of the dynamical matrix for a system, allowing us to measure bend, twist, and stretching modes in dimers of DNA-origami particles for the first time. By measuring the fluctuations of dimers of subunits, we rationalized experimental distributions of self-limited assemblies using elastic energy models of various levels of complexity. This series of comparisons highlighted the critical role that cooperative motions of many subunits play in controlling self-assembly. Lastly, we showed how the insights that one can extract from our approach can be used to rationally redesign colloidal subunits to accurately target a specific assembly outcome. Employing such a closed-loop data-driven design cycle is a crucial part of designing self-closing assemblies, which are prone to polydispersity from binding fluctuations, and may prove useful to the design of open, crystalline assemblies~\cite{duque2024limits, liu2024inverse, posnjak2024diamond}.

While measuring mechanical properties is itself a challenge, understanding the origin of the mechanics, as well as how to alter them, is an exciting prospect to tackle in future studies. In a first pass at understanding the magnitude of the stretching and bending moduli that we measure we develop a simple entropic model for the DNA linkages that mediate the interactions between the subunits (see SI Appendix, Section 3). This model predicts a bending modulus that is remarkably close to the experimental value, however, the anticipated stretching modulus is overestimated. In the model, the bending rigidity depends upon the geometrical location of the interacting DNA sticky ends, which suggests that changing their placement on the edge can alter the rigidity. For example, shifting the linking strands by one or two helices from the midplane is expected to increase the bending rigidity by factors of 9 or 25 respectively. Having larger bending rigidity is known to have an important impact on the fidelity of self-closing structures~\cite{videbaek2022tiling, fang2022polymorphic, videbaek2024economical, tyukodi2024magic} and will be an important control knob when creating more complex assemblies. The discrepancy between our entropic model and the stretching modulus points to the need for more detailed investigations. In our system, aspects such as the fraction of incorporated staple strands or the fraction of bound bases in the DNA sticky ends could have important implications for the inter-subunit mechanics. To this end, tools such as oxDNA~\cite{ouldridge2010dna} could be useful for getting more structure-specific models of the single-particle and multi-particle mechanics.

An important factor for having our model predict the correct breadth of the assembly distribution is the rescaling of the fluctuations from the vitrification temperature to the assembly temperature. Our initial, and somewhat naive, approach is to rescale the fluctuations as if they represent an equilibrium distribution at the vitrification temperature. One aspect that this approximation fails to account for is the effects of changes in the viscosity of the solvent during vitrification. Reports on supercooled water show that the viscosity can increase substantially as the temperature decreases~\cite{hallett1963temperature, dehaoui2015viscosity}, which could have an impact on the rotational diffusion timescale and could change the effective temperature at which the fluctuations become frozen out. Additional modeling effort will be needed to accurately account for how thermal fluctuations change in a rapidly cooling liquid. Our approximation also ignores the possibility that the mechanical properties of the structure itself, as well as the molecular connections between bound subunits, may be temperature-dependent. For DNA, cryo-EM measurements of the persistence length of double-stranded DNA give similar values to those taken near room temperature~\cite{bednar1995determination, geggier2011temperature}, hinting that rapid cooling may not have a dramatic impact on DNA-origami objects. 

While we have demonstrated the effectiveness of cryo-EM in helping to predict and understand larger-scale assemblies of DNA-origami subunits~\cite{wei2024economical}, these methods could be further used to model a variety of nanoscale assemblies.  We hope our analysis framework will find use in much broader classes of biomolecular assemblies, for example, de-novo designed proteins, which have recently created increasingly complex self-assembled structures~\cite{shen2018novo, li2023accurate, lutz2023top, pillai2024novo}. Similarly, despite some measurements of the mechanical properties of natural biological assemblies, measuring the mechanical properties of native subunits would enable modeling cellular processes from a more fundamental level. Aside from predicting assembly outcomes, these methods enable quantitative prediction, and ultimately, engineering of the collective mechanical response of the assembled structures itself, for example, the flexural stiffness or axial compressive stiffness of tubules. Such studies could be validated using AFM nanoindentation experiments~\cite{zhou20203d, li2022mechanical}. Taken together, these considerations point to new opportunities to directly engineer and measure the collective mechanics of self-assembling structures, allowing one to directly link the design to the pair-wise quantification and ultimately to the materials properties.

\begin{acknowledgments}
We acknowledge Hendrik Dietz and Fabian Kohler for fruitful initial discussions and for introducing T.E.V. to multi-body refinement. T.E.V. acknowledges Fernando Caballero, Layne B. Frechette, and Naren Sundararajan for helpful discussions about simulations. We thank Berith Isaac and Amanda Tiano for their technical support with electron microscopy. TEM images were prepared and imaged at the Brandeis Electron Microscopy facility. This work is supported by the Brandeis University Materials Research Science and Engineering Center (MRSEC) (NSF DMR-2011846). 
\end{acknowledgments}

\textbf{Data and materials availability:} The cryo-EM data from this study have been deposited in the Electron Microscopy Data Bank with the following accession codes: EMD-48633, EMD-48634, EMD-48635, EMD-48636, and EMD-48637. Designs of DNA origami used in this work can be found in the repository Nanobase and are accessible at https://nanobase.org/structure/234 and https://nanobase.org/structure/258.

\bibliography{main.bib}


\end{document}


\title{Supplemental Appendix for  ``Measuring multisubunit mechanics of geometrically-programmed colloidal assemblies via cryo-EM multi-body refinement''}

\author{Thomas E. Videb\ae k}
\email{videbaek@brandeis.edu}
\affiliation{Martin A. Fisher School of Physics, Brandeis University, Waltham, MA, 02453, USA}
\author{Daichi Hayakawa}
\affiliation{Martin A. Fisher School of Physics, Brandeis University, Waltham, MA, 02453, USA}
\author{Michael F. Hagan}
\affiliation{Martin A. Fisher School of Physics, Brandeis University, Waltham, MA, 02453, USA}
\author{Gregory M. Grason}
\affiliation{Department of Polymer Science and Engineering, University of Massachusetts, Amherst, Massachusetts 01003, USA}
\author{Seth Fraden}
\affiliation{Martin A. Fisher School of Physics, Brandeis University, Waltham, MA, 02453, USA}
\author{W. Benjamin Rogers}
\email{wrogers@brandeis.edu}
\affiliation{Martin A. Fisher School of Physics, Brandeis University, Waltham, MA, 02453, USA}

\maketitle

\setcounter{figure}{0}
\makeatletter 
\renewcommand{\thefigure}{S\arabic{figure}}

\section*{Materials and Methods}

\textbf{Folding DNA origami.} To assemble our DNA origami monomers, we make a solution with 50 nM of p8064 scaffold (Tilibit), 200 nM of each staple strand (Integrated DNA Technologies [IDT]; Nanobase structures 234 and 258~\cite{Nanobase} for sequences), and 1x folding buffer. We then anneal this solution using a temperature protocol described below. Our folding buffer, from here on referred to as FoBX, contains 5~mM  Tris Base, 1~mM EDTA, 5~mM NaCl, and X~mM MgCl$_2$. We use a Tetrad (Bio-Rad) thermocycler to anneal our samples.

To find the best folding conditions for each sample, we follow a standard screening procedure to search multiple MgCl$_2$ concentrations and temperature ranges~\cite{Wagenbauer2017,Hayakawa2022}, and select the protocol that optimizes the yield of monomers while limiting the number of aggregates that form. All particles used in this study were folded at 17.5~mM MgCl$_2$ with the following annealing protocol: (i) hold the sample at 65~$^\circ$C for 15 minutes,  (ii) ramp the temperature from 58~$^\circ$C to 50~$^\circ$C with steps of 1~$^\circ$C per hour, (iii) hold at 50~$^\circ$C until the sample can be removed for further processing. 

\textbf{Agarose gel electrophoresis.} We use agarose gel electrophoresis to assess the folding protocols and purify our samples with gel extraction. We prepare all gels by bringing a solution of 1.5\%~(w/w) agarose in 0.5X TBE to a boil in a microwave. Once the solution is homogenous, we cool it to 60~$^\circ$C using a water bath. We then add MgCl$_2$ and SYBR-safe (Invitrogen) to have concentrations of 5.5~mM MgCl$_2$ and 0.5x SYBR-safe. We pour the solution into an Owl B2 gel cast and add gel combs (20 $\upmu$L wells for screening folding conditions or 200~$\upmu$L wells for gel extraction), which cools to room temperature. A buffer solution of 0.5x TBE and 5.5~mM MgCl$_2$, chilled at 4~$^\circ$C for an hour, is poured into the gel box. Agarose gel electrophoresis is run at 110 V for 1.5--2 hours in a 4~$^\circ$C cold room. We scan the gel with a Typhoon FLA 9500 laser scanner (GE Healthcare) at 100~$\upmu$m resolution. 

\textbf{Sample purification.} After folding, we purify our DNA origami particles to remove all excess staples and misfolded aggregates using gel purification. The folded particles are run through an agarose gel (now at a 1xSYBR-safe concentration for visualization) using a custom gel comb, which can hold around 2~mL of solution per gel. We use a blue fluorescent light table to identify the gel band containing the monomers. The monomer band is then extracted using a razor blade. We place the gel slices into a Freeze ’N Squeeze spin column (Bio-Rad), freeze it in a --20~$^\circ$C freezer for 5 minutes, and then spin the solution down for 5~minutes at 12~krcf. The concentration of the DNA origami particles in the subnatant is measured using a Nanodrop (Thermo Scientific). We assume that the solution consists only of monomers, where each monomer has 8064 base pairs.

Since the concentration of particles obtained after gel purification is typically not high enough for assembly, we concentrate the solution using ultrafiltration~\cite{Wagenbauer2017}. First, a 0.5-mL Amicon 100-kDa ultrafiltration spin column (Millipore) is equilibrated by centrifuging down 0.5~mL of 1xFoB5 buffer at 5~krcf for 7~minutes. Then, the DNA origami solution is added and centrifuged at 14 krcf for 15 minutes. We remove the flow-through and repeat the process until all of the DNA origami solution is filtered. Finally, we flip the filter upside down into a new Amicon tube and spin down the solution at 1~krcf for 2~minutes. The concentration of the final DNA origami solution is then measured using a Nanodrop.

\textbf{Tubule assembly.} Assembly experiments are conducted with DNA origami particle concentrations ranging from 2~nM to 30~nM. For assemblies that are made up of multiple colors, the quoted concentration is the total concentration of all subunits, e.g. for a 10-nM experiment with $N$ colors, each color has a concentration of 10/$N$ nM. Assembly solutions have volumes up to 50~$\upmu$L with the desired DNA origami concentration in a 1xFoB20 buffer. The solution is placed in a 200~$\upmu$L PCR tube and loaded into a thermocycler (Bio-Rad), which is held at a constant temperature, ranging between 30 $^\circ$C and 50~$^\circ$C. The thermocycler lid is held at 100~$^\circ$C to prevent condensation of water on the cap of the PCR tube.

Since DNA hybridization is highly sensitive to temperature, we expect that there should be a narrow range of temperatures over which the system can assemble by monomer addition. To make sure that we assemble tubules within this regime, we prepare many samples over a broad range of temperatures. At high temperatures, we find that there are no large assemblies, implying that we are above the melting transition for our ssDNA interactions. As we lower the temperature, we find a transition to the formation of assembled tubules, but with increasing defect density as the temperature decreases. For this reason, all assembly experiments are conducted just below the melting transition. 

\textbf{Negative-stain TEM.} We first prepare a solution of uranyl formate (UFo). We boil doubly distilled water to deoxygenate it and then mix in UFo powder to create a 2\%~(w/w) UFo solution. We cover the solution with aluminum foil to avoid light exposure and vortex it vigorously for 20~minutes, after which we filter the solution with a 0.2~$\upmu$m filter. Lastly, we divide the solution into 0.2~mL aliquots, which are stored in a -80 $^\circ$C freezer until further use.

Before each negative-stain TEM experiment, we take a 0.2~mL UFo aliquot out from the freezer to thaw at room temperature. We add 4~$\upmu$L of 1~M NaOH and vortex the solution vigorously for 15~seconds. The solution is centrifuged at 4~$^\circ$C and 16~krcf for 8~minutes. We extract 170~$\upmu$L of the supernatant for staining and discard the rest.

The EM samples are prepared using FCF400-Cu grids (Electron Microscopy Sciences). We glow discharge the grid prior to use at -20~mA for 30 seconds at 0.1 mbar, using a Quorum Emitech K100X glow discharger. We place 4 $\upmu$L of the sample on the carbon side of the grid for 1 minute to allow adsorption of the sample to the grid. During this time, 5 $\upmu$L and 18 $\upmu$L droplets of UFo solution are placed on a piece of parafilm. After the adsorption period, the remaining sample solution is blotted on 11 $\upmu$m Whatman filter paper. We then touch the carbon side of the grid to the 5 $\upmu$L drop and blot it away immediately to wash away any buffer solution from the grid. This step is followed by picking up the 18 $\upmu$L UFo drop onto the carbon side of the grid and letting it rest for 30 seconds to deposit the stain. The UFo solution is then blotted and any excess fluid is vacuumed away. Grids are allowed to dry for a minimum of 15 minutes before insertion into the TEM. 

We image the grids using an FEI Morgagni TEM operated at 80 kV with a Nanosprint5 CMOS camera (AMT). The microscope is operated at 80 kV and images are acquired between x8,000 to x20,000 magnification. 

\textbf{Cryo-electron microscopy.} Higher concentrations of DNA origami are used for cryo-EM grids than for assembly experiments. To prepare samples, we fold between 1--2 mL of the folding mixture (50 nM scaffold concentration), gel purify it, and concentrate the sample by ultrafiltration, as described above. EM samples are prepared on glow-discharged C-flat 1.2/1.3 400 mesh grids (Protochip). To ensure that dimers have formed before plunging, monomers with a single active side were mixed 1:1 with 1xFoB35, bringing their salt concentration to 20~mM MgCl$_2$. The solution then sits at room temperature for 30~min. Plunge-freezing of grids in liquid ethane is performed with an FEI Vitrobot VI with sample volumes of 3 $\upmu$L, wait time of 60 s, blot time of 9 s, and a blot force of 0 at 22 $^\circ$C and 100\% humidity. 

Due to microscope time limitations, Tecnai F30 and F20 TEMs were used in this study. The Tecnai F30 TEM is equipped with a field emission gun electron source operated at 300 kV, a Compustage, and an FEI Falcon II direct electron detector. F30 images were taken at a magnification of x39000 with a pixel size of 2.87 Angstrom. The Tecnai F20 TEM is equipped with a field emission gun electron source operated at 200 kV, a Compustage, and a Gatan Oneview CMOS camera. F20 images were taken at a magnification of either x29000 with a pixel size of 3.757 Angstrom (for dimers) or x50000 with a pixel size of 2.231 Angstrom (for the twist-corrected monomer). For both machines, particle acquisition is performed with SerialEM. The defocus is set between -1.5 and -4 $\upmu$m for all acquisitions. The side 1 dimer was imaged on the F30 while all other structures, the side 2 dimer, the side 3 dimer, the hexamer, and the twist-corrected monomer, were imaged on the F20.

\textbf{Single-particle reconstruction.} \noindent 
Image processing is performed using RELION-4~\cite{kimanius2021RELION4}. Contrast-transfer-function (CTF) estimation is performed using CTFFIND4.1~\cite{Rohou2015ctffind}. After picking single particles, we perform a reference-free 2D classification from which the best 2D class averages are selected for processing, estimated by visual inspection. The particles in these 2D class averages are used to calculate an initial 3D model. A single round of 3D classification is used to remove heterogeneous monomers and the remaining particles are used for 3D auto-refinement and post-processing. The post-processed maps are deposited in the Electron Microscopy Data Bank. 

\textbf{Multi-body analysis.} \noindent
Fluctuations of subunits were processed using RELION-4's~\cite{kimanius2021RELION4} multi-body refinement~\cite{nakane2018characterisation}. After getting a postprocessed reconstruction of a dimer using single-particle reconstruction, we create masks around the two triangular bodies using the eraser tool in ChimeraX~\cite{goddard2018ucsf}. These were used in the ``3D multi-body" job in RELION 4 to get the set of fluctuations in translation and rotation of the two bodies in the dimer.

\clearpage

\section{Getting 3D coordinates from RELION's multi-body output} \label{sec:SI-RELIONcoords}

To get the coordinates for any particular particle image it is important to understand how to use the output files from RELION's multi-body job to transform each rigid body to any particular fluctuation state. The outputs from RELION are:
\begin{enumerate}
    \item \code{run\_bodyXYZ.mrc} - refined body density maps
    \item \code{run\_bodyXYZ\_mask.mrc} - mask for the refined body
    \item \code{analyse\_eigenvectors.dat} - eigenvectors for the PCA modes
    \item \code{analyse\_projection\_along\_eigenvectors\_all\_particles.txt} - eigenvalues for each PCA eigenvector for each particle image in the analysis. This is generated with the \code{--write\_pca\_projections} flag.
    \item \code{analyse\_pca\_weights.dat} - normalization coefficients for each real-space coordinate
    \item \code{analyse\_means.dat} -  average value for each real-space coordinate used in the PCA analysis. Note: we needed to add code to produce this output file.
\end{enumerate}

First, we need to convert the transformation coordinates from the eigenvector components of the PCA from RELION to a set of translations and rotations for each body. This follows from a simple expression
\begin{equation}
    \Vec{u}_\mathrm{real}=\Vec{u}_\mathrm{PCA} \textbf{E}_\mathrm{PCA} + \langle \Vec{u} \rangle_\mathrm{real}
\end{equation}
where $\Vec{u}_\mathrm{real}$ is the vector of real-space transformation coordinates, $\mathbf{E}_\mathrm{PCA}$ is the matrix of eigenvectors for the PCA modes, and $\langle \Vec{u} \rangle_\mathrm{real}$ is the mean vector of $\Vec{u}_\mathrm{real}$ averaged over all observed particles used in the analysis. To be explicit, 
\begin{eqnarray}
    \Vec{u}_\mathrm{real} &=& (\alpha_1, \beta_1, \gamma_1, x_1, y_1, z_1, ..., \alpha_{N_\mathrm{b}}, \beta_{N_\mathrm{b}}, \gamma_{N_\mathrm{b}}, x_{N_\mathrm{b}}, y_{N_\mathrm{b}}, z_{N_\mathrm{b}},) \\
    \textbf{E}_\mathrm{PCA} &=& 
    \begin{pmatrix}
        \Vec{e}_1 \\
        \Vec{e}_2 \\
        ... \\
        \Vec{e}_{6N_\mathrm{b}}
    \end{pmatrix} \\
    \Vec{u}_\mathrm{PCA} &=& (\lambda_1, \lambda_2, ..., \lambda_{6N_\mathrm{b}})
\end{eqnarray}
$\alpha,\ \beta,$ and $\gamma$ are Euler angles of rotation and $x,\ y,$ and $z$ are displacements, one for each rigid body up to $N_\mathrm{b}$ bodies. $\Vec{e}_i$ are the PCA eigenvectors for each mode, there will be six times the number of bodies of these, and the $\lambda_i$ are the associated eigenvalues for a particular observation, i.e. for every particle in the multi-body analysis there is a set of $\lambda_1$ to $\lambda_{6N_\mathrm{b}}$ that describe its particular state. Values of $\Vec{u}_\mathrm{real}$ come from the \code{analyse\_projection\_along\_eigenvectors\_all\_particles.txt} file, $\textbf{E}_\mathrm{PCA}$ comes from the \code{analyse\_eigenvectors.dat} file, and $\langle \Vec{u} \rangle_\mathrm{real}$ comes from the \code{analyse\_means.dat} file.

With the transformation coordinates in hand, we can now go about the coordinates of the bodies for a particular fluctuation. To do this we will need to construct transformation matrices that take the average reconstruction of a body to the location of a particular fluctuation using the values of $\Vec{u}_\mathrm{real}$ that the RELION multi-body job output. The steps to do this are as follows:
\begin{enumerate}
    \item Correct $\Vec{u}_\mathrm{real}$ with the output PCA weights found by RELION 
    \item Calculate a rotation matrix and translation vector for each body
    \item Apply the transformations to the bodies
\end{enumerate}
While all of these steps can be decerned from the RELION documentation we will write them down explicitly here.

1. Correcting the transformation parameters with the weights is quite straightforward. In the \code{analyse\_pca\_weights.dat} file, we are provided with weights for each rotation angle and for translations, each direction having the same weight, for each body. This gives a weight vector of the form
\begin{equation}
    \Vec{w} = (w_{\alpha,1}, w_{\beta,1}, w_{\gamma,1}, w_{t,1}, w_{t,1}, w_{t,1}, ... w_{\alpha,N_\mathrm{b}}, w_{\beta,N_\mathrm{b}}, w_{\gamma,N_\mathrm{b}}, w_{t,N_\mathrm{b}}, w_{t,N_\mathrm{b}}, w_{t,N_\mathrm{b}})
\end{equation}
where $w_{\alpha,i},\ w_{\beta,i},\ w_{\gamma,i}$ and $w_{t,i}$ correspond to the \code{body}, \code{rot}, \code{tilt}, and \code{offset} columns for body $i$ in the \code{analyse\_pca\_weights.dat} file. This vector is used to do a normalization correction by
\begin{equation}
    \Vec{u}'_\mathrm{real} = \Vec{u}_\mathrm{real}/\Vec{w}
\end{equation}

2. The translation vector is also straightforward, it is constructed from the values of $x'_i$, $y'_i$, $z'_i$ in $\Vec{u}'_\mathrm{real}$ such that for body $i$ the translation vector is 
\begin{equation}
\Vec{T}_i = (x'_i, y'_i, z'_i)
\end{equation}
The rotation matrix is a bit more involved and takes the form 
\begin{equation}
    \mathbf{A}_\mathrm{T} = \mathbf{O}^T \mathbf{A}_{90} \mathbf{A}_\mathrm{resi} \mathbf{O}
\end{equation}
where $\mathbf{O}$ is an orientation matrix, $\mathbf{A}_\mathrm{resi}$ is a residual transformation matrix that comes from the Euler angles in $\Vec{u}_\mathrm{real}$, and $\mathbf{A}_{90}$ is a -90 degree rotation about the $y$-axis.

The orientation matrix, $\mathbf{O}$, is different for each body in the analysis. Each body, when setting up the multi-body refinement job, is told which body it rotates relative to. The orientation matrix is the rotation matrix that aligns the axis vector between a body and its relative body to the $z$-axis. For instance, if body $i$ is specified relative to body $j$, then the axis vector would be $\Vec{a}=\Vec{C}_j - \Vec{C}_i$, where $\Vec{C}_i$ is the vector pointing to the center of mass of body $i$. Then the orientation matrix is
\begin{equation}
    \mathbf{O} = \begin{pmatrix}
        |\vec{a}| & -a_x a_y / |\vec{a}| & -a_x a_z / |\vec{a}| \\
        0 & a_z / |\vec{a}| & -a_y / |\vec{a}| \\
        a_x & a_y & a_z
    \end{pmatrix}
\end{equation}
which we take directly from the \code{transformations.cpp} code in the RELION 4 Github page. The center of mass coordinates come from the output body densities maps (from \code{run\_bodyXYZ.mrc}) of the multi-body job after applying the mask for the body (from \code{run\_bodyXYZ\_mask.mrc}).

The residual rotation matrix, $\mathbf{A}_\mathrm{resi}$, comes from the Euler angles, $\alpha$, $\beta$, and $\gamma$ in $\vec{u}'_\mathrm{real}$ for the body of interest. RELION uses the $Z_\gamma$-$Y_\beta$-$Z_\alpha$ convention for Euler angles to construct the rotation matrix but with negative angles. The residual rotation matrix is then
\begin{equation}
    \mathbf{A}_\mathrm{resi} = \begin{pmatrix}
        \cos \gamma \cos \beta \cos \alpha - \sin \gamma \sin \alpha & \cos \gamma \cos \beta \sin \alpha + \sin \gamma \cos \alpha & -\cos \gamma \sin \beta \\
        -\sin \gamma \cos \beta \cos \alpha - \cos \gamma \sin \alpha & -\sin \gamma \cos \beta \sin \alpha + \cos \gamma \cos \alpha & \sin \gamma \sin \beta \\
        \sin \beta \cos \alpha & \sin \beta \sin \alpha & \cos \beta
    \end{pmatrix}
\end{equation}
which we take directly from the \code{euler.cpp} code on the RELION 4 Github page.

3. We can now apply the rotations and translations to the bodies. Note that these transformations are applied to the pixel coordinates of the MRC file for the density maps. For a given coordinate in a body of interest, $\vec{v}$, we perform the following steps in order
\begin{eqnarray}
\vec{v} &\rightarrow& \vec{v} - \vec{C}_\mathrm{BOX} \\
\vec{v} &\rightarrow& \mathbf{A}_\mathrm{T} \vec{v} \\
\vec{v} &\rightarrow& \vec{v} + \vec{C}_\mathrm{BOX} \\
\vec{v} &\rightarrow& \vec{v} + \vec{T}
\end{eqnarray}
where $\vec{C}_\mathrm{BOX}$ is the vector pointing to the center of the box of the MRC file.

\subsection{Check that transformations give back the PCA movies from multi-body.}

\begin{figure*}[!ht]
    \centering
    \includegraphics[width=\linewidth]{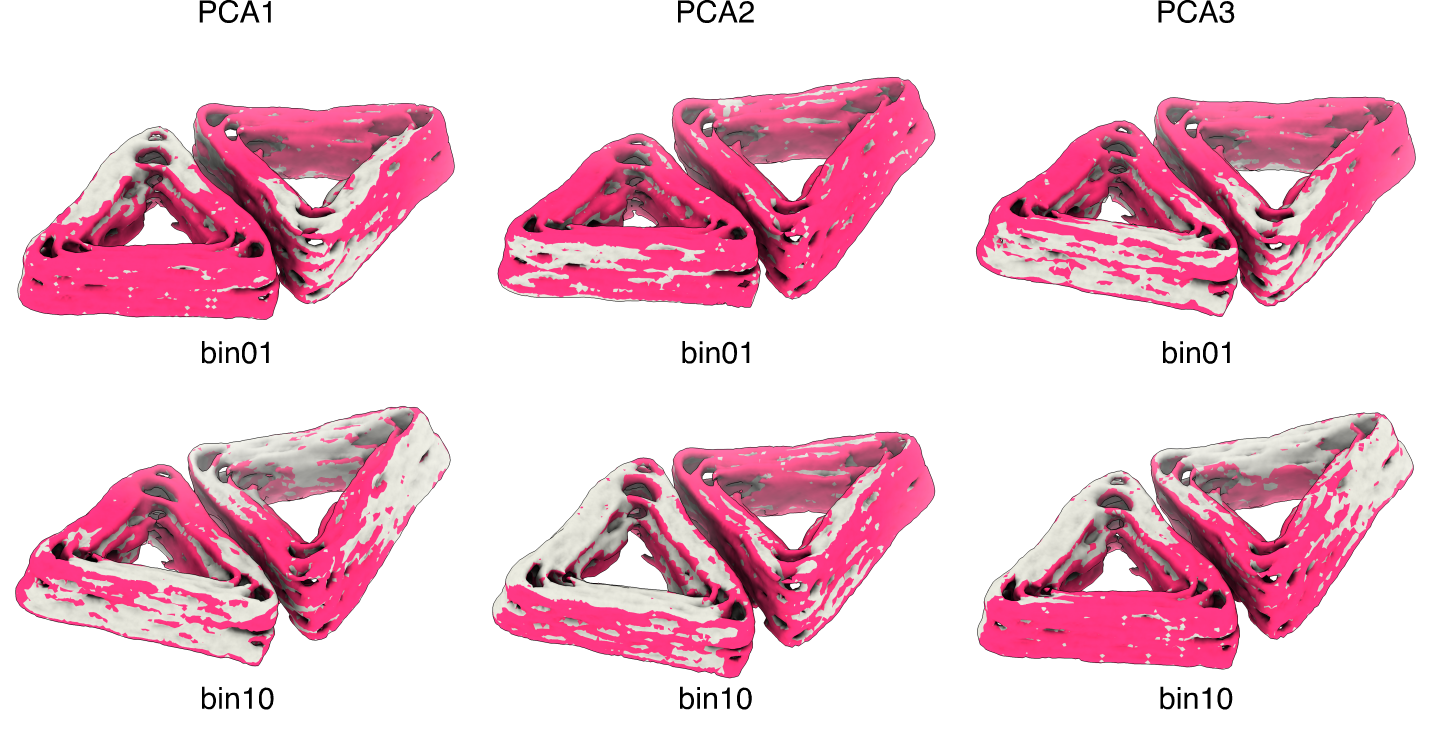}
    \caption{\textbf{Analysis transformation confirmation}. Grey bodies are surfaces of density maps from RELION 4's PCA movies and red bodies are surfaces of density maps generated from transforming the refined body maps using the generated PCA eigenvectors and eigenvalues. Maps are from the side 1 dimer for the tubule.}
    \label{Fig:sfig-pcacheck}
\end{figure*}

To confirm that our manipulation of the bodies is consistent with the outputs provided by RELION for the PCA modes, we recreate the multi-body configurations for the 10th and 90th percentile bins for the first three PCA modes for the Side 1 dimer (Fig.~\ref{Fig:sfig-pcacheck}). We find that the density map surfaces are visible within each other and that we can recreate the RELION-generated density maps for the PCA modes within a fraction of a voxel.

\subsection{Additional code added to RELION scripts}

To get the \code{analyse\_mean.dat} output file we have added the following code after line 264 in the \code{flex\_analyser.cpp} code of RELION.

\begin{verbatim}

		// Write the mean file data for particles to a file before doing PCA on them
		FileName fn_mean = fn_out + "_means.dat";
		std::ofstream f_mean(fn_mean);
		std::cout << " Means (rotations only):" << std::endl;
		for (int j = 0; j < means.size(); j++)
		{
			std::string stro = "";
			if (j % 6 == 0)
				stro = "rot";
			else if (j % 6 == 1)
				stro = "tilt";
			else if (j % 6 == 2)
				stro = "psi";
			else if (j % 6 == 3)
				stro = "x";
			else if (j % 6 ==  4)
				stro = "y";
			else if (j % 6 == 5)
				stro = "z";
			if (stro != "")
			{
				stro +=  "-body-" + integerToString(1 + (j / 6));
				f_mean << stro << " ";
				if (j % 6 < 3) {
					std::cout << std::setw(12) << std::right << std::fixed;
					std::cout << stro;
				}    }    }
		std::cout << std::endl;
		f_mean << std::endl;
		std::cout << " Full means including translations are written to " 
  << fn_mean << std::endl;

		f_mean << std::scientific;

			for (int j =0; j < means.size(); j++)
			{
				if (j > 0) f_mean << " ";
				f_mean << means[j];
			}
			f_mean << std::endl;

		f_mean.close();

\end{verbatim}

\clearpage

\section{Calculating relative coordinates for dimers}

\begin{figure*}[!bh]
    \centering
    \includegraphics[width=\linewidth]{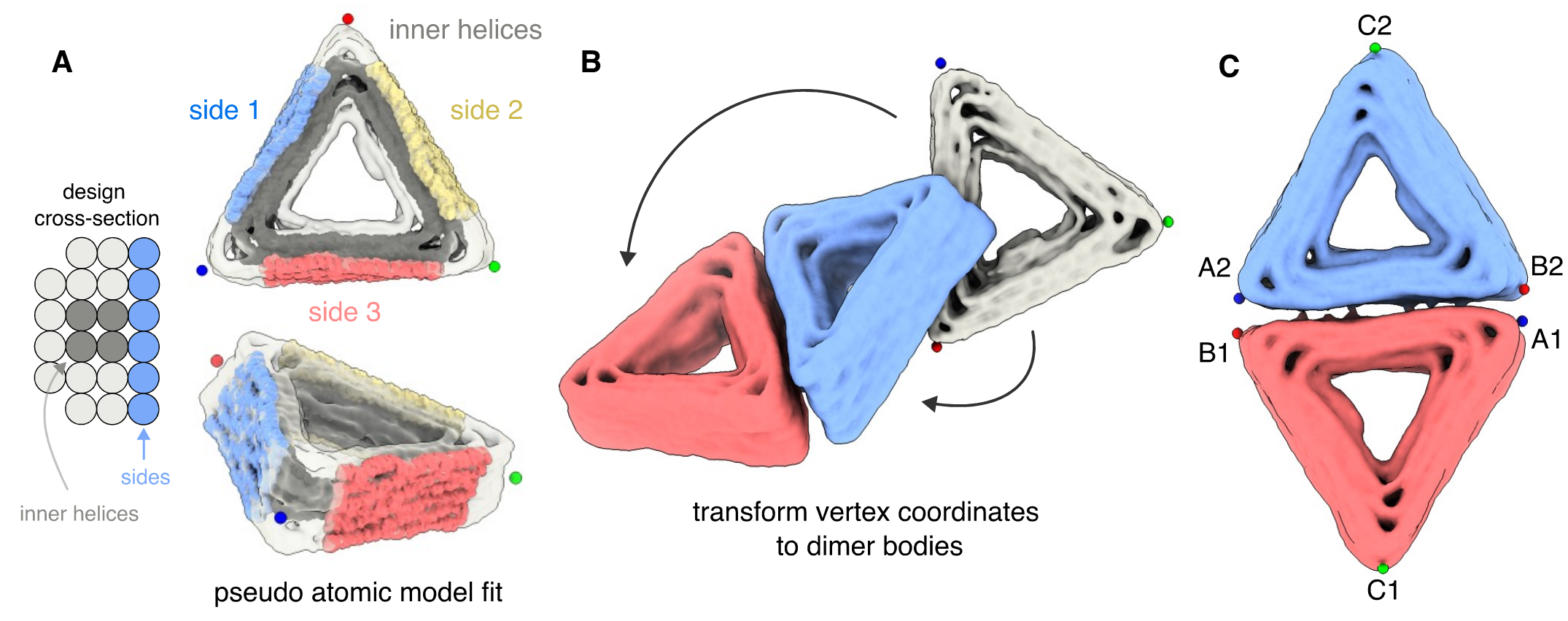}
    \caption{\textbf{Defining coordinates for dimers}. (A) A pseudo-atomic model of the monomer is used to extract the coordinates of helices from the three sides as well as the center of all sides. A cross-section of the design shows which helices are chosen for sides versus the inner helices. These coordinates are used to find points (blue, green, and red spheres) that define the vertices of the triangle. (B) The monomer map is then fit into either of the dimer particles to get the transformation needed to place vertex markers on the two dimer particles. (C) Once the markers are in place, we label them A, B, or C and use their relative positions to calculate the coordinate systems and relative coordinates shown in Fig.~2D in the main text.}
    \label{Fig:sfig-dimercoord}
\end{figure*}

As shown in Fig.~2D in the main text, we define new coordinate systems for both bodies in our dimer reconstructions so that we can make measurements of the relative angles and displacements between the two bodies. To do this, we first need to choose a consistent set of points based on the reconstructions. We choose to use the vertices of the triangles as these points. We fit a pseudo-atomic model (PAM) to a reconstruction of the monomer using the methods in ref.~\cite{kube2020revealing} implemented in https://github.com/elija-feigl/dnaFit. In brief, this method (i) takes the design file from caDNAno and simulates an atomic model of the structure using mrdna~\cite{maffeo2020mrdna}, it then (ii) uses the electron density of the cryo-EM reconstruction as a potential to relax the atomic model into a position that best fits the cryo-EM map. This allows one to choose helices and base pairs from the design and find their coordinates in the cryo-EM map. Once we have the PAM we can select coordinates from our cryo-EM electron density map based on the initial DNA origami design. We extract coordinates corresponding to the outer layer of helices for each of the three sides as well as the inner four helices of the structure. We fit these all to planes and use the plane intersections to define the vertices of the triangle (Fig.~\ref{Fig:sfig-dimercoord}A).

To get the initial vertex positions for the dimers we need to map the vertex positions that we fit on the monomer to each triangle in the dimer (Fig.~\ref{Fig:sfig-dimercoord}B). Using built-in commands in ChimeraX makes this an easy task. First, we fit the monomer map in one of the dimer triangles using the \code{Fit in Map} command. Then we use the \code{measure rotation} command to get both a rotation matrix, $\mathbf{A}$, as well as a translation vector, $\vec{T}$. With these, we can transform the vertex positions we got from the monomer fitting to the dimer by $\vec{v}_\mathrm{dimer} = \mathbf{A}\vec{v}_\mathrm{mono} + \vec{T}$, where $\vec{v}$ is a vertex position. Once we have the coordinates for the vertices for both dimers we can use the transformations described in SM Sec.~\ref{sec:SI-RELIONcoords} to get vertex positions for any fluctuation for a dimer (Fig.~\ref{Fig:sfig-dimercoord}C).

The coordinate system for a triangle with vertex positions of $A$, $B$, and $C$ are given by 
\begin{eqnarray}
    \hat{y} &=& (\vec{OB} - \vec{OA})/\mathrm{norm}(\vec{OB} - \vec{OA}) \\
    \hat{z} &=& (\vec{OB} \times \vec{OC})/\mathrm{norm}(\vec{OB} \times \vec{OC}) \\
    \hat{x} &=& \hat{y} \times \hat{z}
\end{eqnarray}
where $\vec{OA}$ is the vector from the origin pointing to position $A$. Once we have coordinate systems for each particle in the dimer we can compute the relative angles between the bodies, $\theta_\mathrm{R}$, $\theta_\mathrm{T}$, and $\theta_\mathrm{B}$. These are calculated by projecting a coordinate direction for one of the bodies onto a corresponding plane of coordinates for the other body and then finding the angle between the projected vector and a coordinate vector. For example, to get the bending angle, $\theta_\mathrm{B}$, we project the $z$-axis of body 2, $\hat{z}_2$, onto the $x-z$ plane of body 1. We then measure the angle between the projected $\hat{z}_2$ and the $x$-axis of body 1 and subtract $\pi/2$ (this is done instead of taking the angle between the two $z$-axis directions to avoid numerical problems of measuring the angle between parallel vectors). 

We first define a projection function, $P$, and an angle function, $\Theta$, as follows
\begin{eqnarray}
    P(\vec{a}, \vec{b}) &=& (\vec{a}-(\vec{a}\cdot\vec{b})\vec{b})/\mathrm{norm}(\vec{a}-(\vec{a}\cdot\vec{b})\vec{b}) \\
    \Theta(\vec{a}, \vec{b}) &=& \arctan ( \mathrm{norm}( \vec{a}\times\vec{b}) /(\vec{a}\cdot \vec{b}) )
\end{eqnarray}
With these, we compute the relative angles between the bodies as
\begin{eqnarray}
    \theta_\mathrm{R} = \Theta(\hat{y}_1, P(\hat{x}_2, \hat{z}_1)) -\pi/2 \\
    \theta_\mathrm{T} = \Theta(\hat{z}_1, P(\hat{y}_2, \hat{x}_1)) -\pi/2 \\
    \theta_\mathrm{B} = \Theta(\hat{x}_1, P(\hat{z}_2, \hat{y}_1)) -\pi/2
\end{eqnarray}
with subscripts 1 and 2 denoting coordinate vectors for bodies 1 and 2 respectively.

We also compute the relative displacements between the two bodies, specifically the displacement with respect to the interacting edges. First, we find the mid-point of the edge as 
\begin{eqnarray}
 \vec{E} = (\vec{OA} + \vec{OB})/2    
\end{eqnarray}
Then we take the vector between these edge points and take its components in the reference frame of body 1 as the relative displacements
\begin{eqnarray}
dx = (\vec{E}_2 - \vec{E}_1)\cdot \hat{x}_1 \\
dy = (\vec{E}_2 - \vec{E}_1)\cdot \hat{y}_1 \\
dz = (\vec{E}_2 - \vec{E}_1)\cdot \hat{z}_1 \\    
\end{eqnarray}

\section{Entropic model for DNA linkers}

To elucidate the factors that control the inter-subunit mechanics, we developed a simple model to estimate the stretching and bending moduli. We assume that stretching and bending are opposed by the deformation-free energy of the $N_\mathrm{linker}=6$ DNA linker molecules at each interface. The linkers are affixed to two neighboring DNA helices on each subunit interface (see Fig.~2A in the main text), with the two helices separated by a vertical distance $h=2.6$ nm. Each linker contains a double-stranded (DS) region with $N_\mathrm{bp}=6$ base pairs and 2 nucleotides of single-stranded DNA (ssDNA) at either end. The total stretching modulus for the subunit-subunit interface is given by $U_\mathrm{stretch} = N_\mathrm{linker}k_1(x-x_0)^2$, with $x$ the displacement, $x_0$ the equilibrium displacement, and $k_1$ the stretching modulus for one linker. We assume, and subsequently confirm, that the linker flexibility is dominated by the ssDNA regions since the DS region has a modulus that is two orders of magnitude larger. Experiments on stretching ssDNA under force have not led to a consensus estimate of the modulus. 
Therefore, to estimate the stretching modulus, we use the well-known formula for the entropic cost for extension of a flexible polymer, $k_1 =\ 3 k_\mathrm{B}T/N_\mathrm{k}l_\mathrm{k}^2$, with $l_\mathrm{k}$ as the Kuhn length (of the statistical segment length) and $N_\mathrm{k}$ the number of statistical segments~\cite{de1990introduction}. For ssDNA at the experimental salt concentration, $l_k\approx 1.5$ nm, approximately two nucleotides~\cite{smith1996overstretching}, and thus each ssDNA region corresponds to roughly 1 segment, so $N_\mathrm{k}=2$. The extension expression is derived for the $N_\mathrm{k}\rightarrow\infty$ limit; while in practice it performs well for polymers as small as $N_\mathrm{k}\approx 5-10$, for our system we can only use it as an order of magnitude estimate. Keeping that in mind, the total stretching modulus for the subunit interface is given by $k_\mathrm{x}\approx 9\ k_\mathrm{B}T/\mathrm{nm^2}$.

To estimate the bending modulus, we assume that the neutral surface (along which the displacement $x$ is independent of the bend deformation) lies between the two helices that the linkers are affixed to. Thus, the linkers are at vertical distances $h/2$ above and below the neutral surface. Assuming that the stretch and bend deformations are uncoupled (as found in the experiments), the bending energy for an angle $\theta$ is given by $U_\mathrm{bend}=B\theta^2$ with $B=N_\mathrm{linker}k_1 h^2 /4$. For our geometry and the above estimate of stretching modulus, we obtain $B\approx 15\ k_\mathrm{B}T$.

\textit{Discussion}. We can make a few remarks about these results. First, although presumably fortuitous given the level of approximation, the predicted bending modulus is remarkably close to the experimental estimate, especially once rescaled to account for the vitrification effects ($B=17\ k_\mathrm{B}T$). 
Due to the geometrical nature for the origin of the bending modulus, the model suggests design strategies; for example, moving the binding interactions by one or two helices further apart would increase the bending modulus by factors of 9 and 25 respectively. Second, the stretching modulus here is larger than the experimental estimate ($k_{\rm x} = 0.5\ k_{\rm B}T/{\rm nm}^2$). One assumption we made above is that all bases in the dsDNA domains are bound for all linkers. The stretching modulus may be more sensitive to unbound portions of strands or missing linking strands than the bending modulus. Despite this, the relatively close agreement with experimental measurements suggest that this simple model captures the relevant forces that control the mechanical properties of the subunit interfaces.

\textit{Comaparison to dsDNA regions.} Finally, to test the assumption that flexibility is dominated by the ssDNA regions, we compute the stretching modulus under the converse – flexibility being dominated by the DS regions. The stretch modulus for dsDNA, defined as $S/L_0=fx$, $L_0$ the unperturbed contour length, and $f$ the applied force is on the order of $S\approx 1000$ on, from force-extension experiments and atomistic simulations~\cite{aggarwal1995structure, garai2015dna, marin2017understanding}. Using $L_0 = 0.24 N_\mathrm{bp}$ we obtain $k_\mathrm{x,ds} = 735\ \mathrm{pn/nm^2}$, thus justifying the assumption.

\section{Elastic energy model for tubules}

\begin{figure*}[!th]
    \centering
    \includegraphics[width=\linewidth]{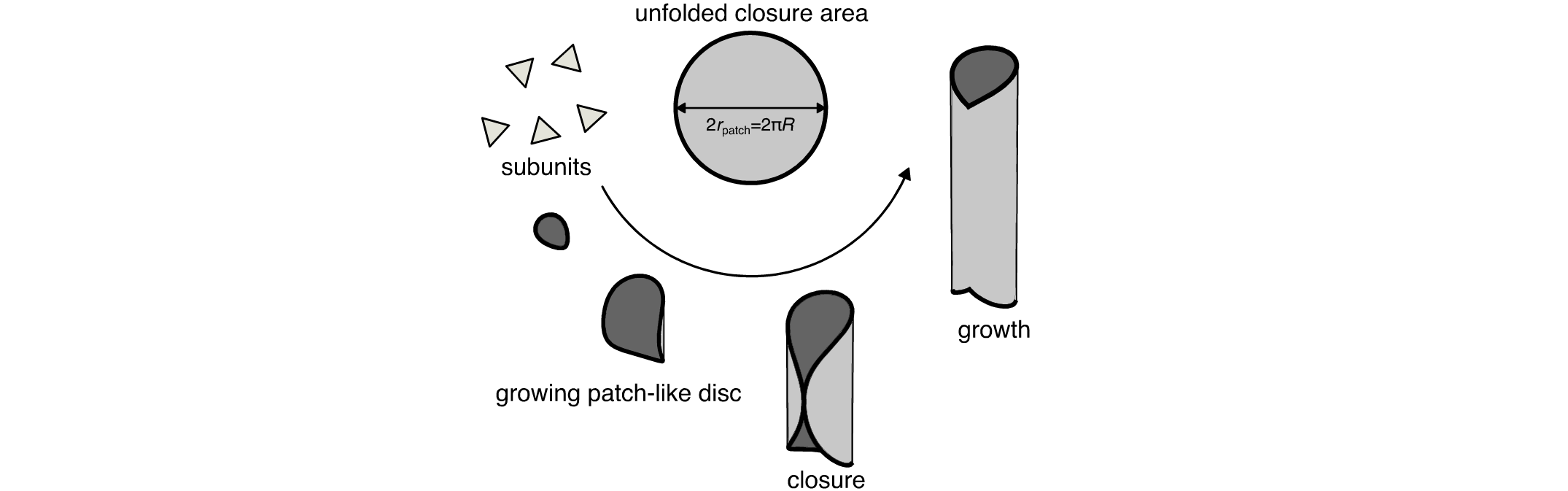}
    \caption{\textbf{Model for tubule formation}. Sketch of tubule growth from a patch-like disc to a closed assembly.}
    \label{Fig:sfig-helfrich}
\end{figure*}

For modeling the distribution of tubules we use an elastic energy model for our tubule assemblies that we have developed in previous work~\cite{videbaek2022tiling,videbaek2024economical,Hayakawa2022,fang2022polymorphic}, which we reproduce here. We assume that the growth of assemblies in the tubule system roughly follows the following path: (i) subunits will initially form a growing patch, which adopts a curvature that is related to the binding angles programmed into the subunits; this curvature is also subject to thermal fluctuations, (ii) if the binding energies for each side of the triangular subunits are similar, we assume that the patch will then grow isotropically until it is large enough to close upon itself, (iii) after closure, the assembly is classified as a tubule and can continue growth by extending (Fig.~\ref{Fig:sfig-helfrich}). From prior work, we assume that the assembly of tubules is slow and that monomers attach to and detach from the growing assemblies so that it can be modeled as quasi-equilibrium (i.e. a fixed chemical potential with free monomers). In our experimental results, by contrast, we assume that tubule type is effectively locked in when the sheet closes on itself because, at this point, disassembly involves breaking multiple bonds and thus has a large activation barrier to change types. This assumption means that the dispersity of states in the assembly can be directly tied back to the near-equilibrium shape fluctuations of the pre-closure sheet.

To determine the distribution of tubule types we estimate of the difference in the free energy between different tubule types at the closure point. The free energy per monomer for a given ($m,n$) tubule, $g^{(m,n)}$, can be approximated by 
\begin{equation}
g^{(m,n)} = - \left( \frac{3}{2}E_\mathrm{bind} + Ts  \right) + E^{(m,n)}_\mathrm{elastic}
\end{equation}
where $-E_\mathrm{bind}$ is the edge binding energy, $T$ is the temperature, $s$ is the per-monomer entropy of bound particles, and $E^{(m,n)}_\mathrm{elastic}$ is the elastic energy (per monomer). Note that we ignore any contribution from the line tension associated with unsatisfied interactions, justified if self-closing patches include have a sufficiently large number of subunits. Given this free energy, the equilibrium probability, $P^{(m,n)}$, of a given tubule type is given by
\begin{equation}
P^{(m,n)} \propto \exp \left(  -N( g^{(m,n)} - \mu )/ k_\mathrm{B}T\right),
\end{equation}
where $N$ is the number of monomers in the assembled patch and $\mu$ is the chemical potential cost to assemble relative to the free monomer state. At equilibrium, the free energy per monomer of the tubule type that minimizes the free energy is approximately equal to the chemical potential, $\mu$~\cite{Hagan2021equilibrium}. Assuming that the entropy per monomer is independent of the tubule type, then $\mu_\mathrm{equilibrium} \lesssim -\left( \frac{3}{2} E_\mathrm{bind} + Ts\right) + E_\mathrm{elastic,min}$, where $E_\mathrm{elastic,min}$ is the residual elastic energy per monomer in the most probable (i.e. minimal elastic energy) tubule type, which results from frustration of pairwise elastic energies of bound monomers in the 2D tubule lattice. Rewriting the elastic energy as $E^{(m,n)}_\mathrm{elastic} = \Delta E^{(m,n)}_\mathrm{elastic} + E_\mathrm{elastic,min}$ lets us rewrite the probability as 
\begin{equation}
P^{(m,n)} \propto \exp \left ( -N \Delta E^{(m,n)}_\mathrm{elastic} /k_\mathrm{B} T \right).
\end{equation}

In a previous simulation study on tubule dynamics~\cite{fang2022polymorphic}, Fang et al. looked at the relationship between the free energy at the closure point and the bending energy for different tubule types. They found that the total free energy difference from the lowest free energy state, $\Delta G^{(m,n)}$, was linearly related to the bending energy difference, $\Delta E^{(m,n)}_\mathrm{bend}$, with a proportionality constant of $0.3 \lesssim \alpha \lesssim 0.5$. To account for this scaling between the free energy and the bending energy, we add a factor of $\alpha$ to our elastic energy term, taking $\alpha=0.3$. This gives a final expression for the tubule-type probability as
\begin{equation}\label{eqn:SI-probability}
P^{(m,n)} \propto \exp \left ( -\alpha N \Delta E^{(m,n)}_\mathrm{elastic} /k_\mathrm{B} T \right).
\end{equation}

The two remaining ingredients are the form of the elastic energy, which we leave to the following subsections, and the number of subunits at closure. Assuming that the growth of the pre-closure assembly is isotropic, we can approximate the closure size as when the diameter of a circular patch matches the circumference of a given tubule type, $2r_\mathrm{patch}=2\pi R$, where $r_\mathrm{patch}$ is the radius of the assembled sheet and $R$ is the radius of curvature of the tubule. Using this relation we get that the assembly has an area of $A=\pi^3 R^2$. Using the area of the closed patch, $A$, we find $N=(4\pi/\sqrt{3})(C/l_0)^2$, where $l_0$ is the edge length of a subunit, and $C$ is the circumference of the tubule. For any ($m,n$) state we can approximate the circumference as $C = \sqrt{m^2 + n^2 + mn}$.

\subsection{Bending energy model}

We begin by considering if only elastic contributions from bending can account for the experimental distribution of tubules we see. We make this simplification since, in the main text, we saw that for all the pairwise measurements the bend angle had the lowest elastic modulus compared to twist and roll angles, as well as a low degree of coupling to other coordinates. The elastic bending energy can be written as
\begin{equation}
E^{(m,n)}_\mathrm{elastic} = \frac{1}{4} \sum_{i \in \mathrm{sides}} B_i (\theta^{(m,n)}_i - \theta_{i,0} )^2
\end{equation}
where $B_i$ is the bending modulus for side $i$, $\theta^{(m,n)}_i$ is the ideal bending angle for side $i$ in a tubule of type ($m,n$), and $\theta_{i,0}$ is the preferred bending angle for side $i$ based on our pairwise measurements.  Notably, this ``bend only'' elastic cost neglects deformations of other modes (e.g. roll and twist) as well as geometric constraints on edge deformations imposed by the 2D bond network between rigid triangles in the lattice of the tubule.  The ideal bending angles for different tubule types can be derived from simple geometric arguments, of which an approachable derivation is shown in Ref.~\cite{Hayakawa2022}. After computing this per monomer elastic energy for all tubule types, we subtract off the lowest energy to get $\Delta E^{(m,n)}_\mathrm{elastic}$ which we can plug into Eqn.~\ref{eqn:SI-probability} to get our model distribution. In the main text, we see that this initial model which only considers bending is not sufficient to explain the origin of the center of the experimental distribution of tubule states that we measure experimentally.

\subsection{Monte Carlo energy minimization}

\begin{figure*}[!th]
    \centering
    \includegraphics[width=\linewidth]{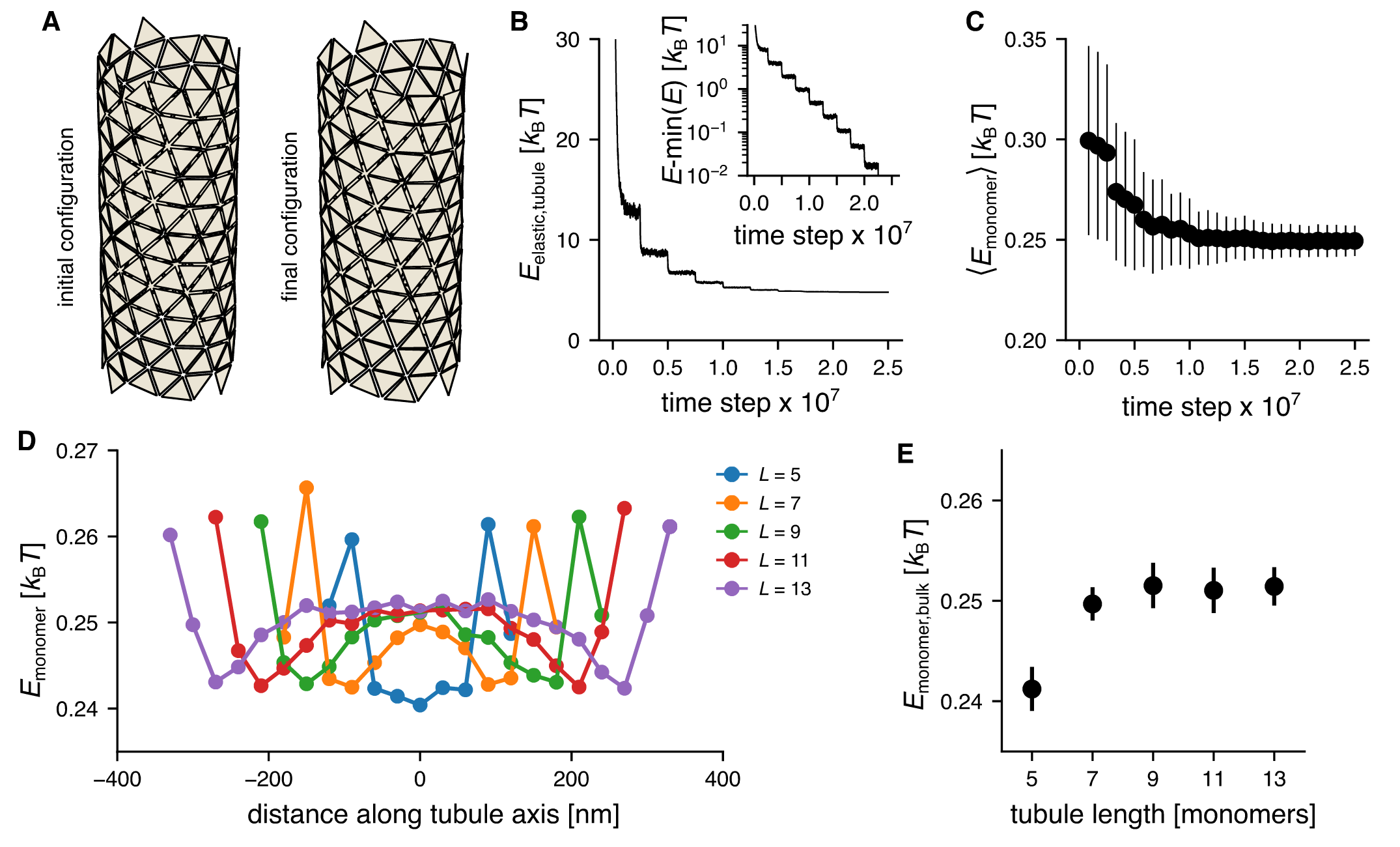}
    \caption{\textbf{Monte Carlo energy minimization of tubules}. (A) The initial configuration and final configurations for a (10,3) tubule with length $L=11$. (B) Plot of the total elastic energy for the configuration during the simulation. The inset shows the difference of the elastic energy and the minimum elastic energy measured during the simulation. (C) Plot of the average elastic energy for monomers neighboring triangles on all three sides. (D) Plot of the monomer elastic energy along the axis of the tubule for different length tubules. Each point is an average over a 25~nm bin. (E) Plot of the monomer elastic energy in the center of different-length tubules. The monomer elastic energy is the average of monomers over a 100~nm segment at the center.}
    \label{Fig:sfig-montecarlo}
\end{figure*}

Since the ``decoupled'' bending model did not capture the observed distributions of tubules, we consider a more complete description that incorporates twist, roll, and stretching degrees of freedom in our model, as well as the interdependence of these due to the 2D network of bonds in the tubule lattice. However, this poses a challenge, since the additional degrees of freedom now make the geometry of any ($m,n$) state not singularly defined, i.e. there are many similar configurations of an ($m,n$) tubule with slight differences in displacements or binding angles. To find the minimum energy configurations for different ($m,n$) tubule states we turn to simulation. In particular, since we have measured the set of preferred relative coordinates and the associated dynamical matrix for each dimer, we want to find the configuration of triangles that form an ($m,n$) state and minimize the elastic energy for each monomer
\begin{equation}
E^{(m,n)}_\mathrm{elastic} = \frac{1}{4}\sum_{i\in\mathrm{sides}}\sum_{\alpha,\beta\in\mathrm{coords}}( q^{(i)}_\alpha - q^{(i)}_{\alpha,0} ) M^{(i)}_{q_\alpha,q_\beta} ( q^{(i)}_\beta - q^{(i)}_{\beta,0} )
\end{equation}
where $q_\alpha$ represents a relative coordinate described in Fig.~2 of the main text and $M^{(i)}$ is the dynamical matrix for side $i$ described in the main text. Once we have the minimum energy configuration, we can compute the average per monomer energy of the tubule and then use that to compute the probability distribution of tubule states using Eqn.~\ref{eqn:SI-probability}.

We use Monte Carlo energy minimization to find the minimum energy configuration for different tubule states. We begin by initializing triangular monomers in a given tubule type with a certain length, SFig.~\ref{Fig:sfig-montecarlo}A shows an example of a (10,3) tubule with a length of 11 monomers. We generate trial moves by giving a small displacement and a random rotational kick to the triangles and using a Metropolis-Hastings criterion to accept the moves, $P_\mathrm{accept} = \mathrm{min}(1, \exp(-\Delta E/k_\mathrm{B}T))$ where $\Delta E$ is the change in elastic energy due to the trial move. After enough moves, the tubule ends up in a lower elastic energy state and has equilibrated.

To equilibrate our tubule configurations, we use an annealing protocol to periodically lower the temperature of our Metropolis-Hastings criterion. The point of lowering the temperature is to get closer to the zero temperature limit, where the actual elastic energy minimum is, while still being able to sample tubule configurations near this minimum. Our simulations involve ten annealing steps where we progressively lower the temperature as well as the range of displacements for trial moves as follows:
\begin{eqnarray} \nonumber
T &\rightarrow& T/2 \\ \nonumber
dX &\rightarrow& dX/\sqrt{2} \\ \nonumber
d\Theta &\rightarrow& d\Theta/\sqrt{2} \nonumber
\end{eqnarray}
where the center of mass displacements occur in a range $[0,\ dX)$ in a random direction and triangle re-orientations occur in an angular range $[0,\ d\Theta)$ about a random axis of rotation. We find that for initial parameters of $k_\mathrm{B}T = 0.01$, $dX = 0.005$ nm, and $d\Theta=0.005$ rad our trial moves have about a 40\% acceptance rate during all of the annealing steps. Using this protocol, we see that the energy of the initial tubule configurations converges (SFig.~\ref{Fig:sfig-montecarlo}B), as does the per monomer elastic energy (SFig.~\ref{Fig:sfig-montecarlo}C).

\begin{figure*}[!th]
    \centering
    \includegraphics[width=\linewidth]{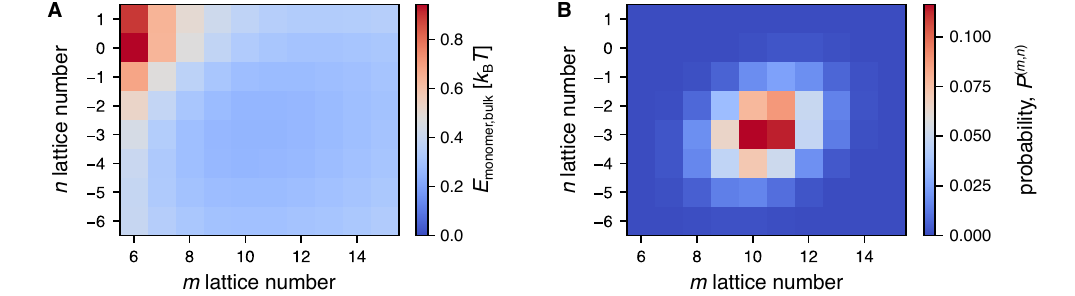}
    \caption{\textbf{Monte Carlo results for varying tubule type}. (A) Monomer energies in the center of tubules of length $L=11$. (B) The probability for different tubule types using the monomer energies shown in (A) and using Eqn.~\ref{eqn:SI-probability}.}
    \label{Fig:sfig-montecarlo-prob}
\end{figure*}

We model the elastic energy of distinct $(m,n)$ states as the per monomer energy of an infinitely long tubule, and thus aim to remove the effects of any elastic relaxation near to free boundary of a finite tubule. Due to the residual elastic stress from the angle frustration in the tubule lattice,  we expect a finite-length relaxed-boundary zone over which edge effects are relevant and past which the monomer is effectively part of an infinite tubule. To see this, we simulate tubule configurations of different lengths and find that for tubules longer than 9 monomers, there is a well-defined monomer energy in the middle of the tubule (SFig.~\ref{Fig:sfig-montecarlo}D and E).  Below this length, all monomers in the tubule feel some boundary effects.

We now compute the bulk monomer energies for different tubule types and use them to calculate the probability of different tubule states. We sample initial tubule configurations for a variety of ($m,n$) tubules with lengths of 11 monomers. Also, since the number of monomers in a tubule of fixed length changes with tubule type, we scale the length of the annealing steps to be proportional to the number of subunits in the tubule. The final configurations are used to calculate the bulk monomer energy for each tubule type (SFig.~\ref{Fig:sfig-montecarlo-prob}A). Note that here we show negative $n$ numbers for the tubules, which corresponds to the lattice direction of side 3 being left-handed along the tubule surface. While we only show the absolute value of $n$ in the main text (due to a limitation of TEM imaging) the prediction of tubules being left-handed helix aligns with TEM tomography that was done in a previous study~\cite{videbaek2024economical}. Finally, we use these calculated monomer energies for the elastic energy in Eqn.~\ref{eqn:SI-probability} and find the model probability distribution for different tubule types (SFig.~\ref{Fig:sfig-montecarlo-prob}B).

\section{Hexamer reconstruction}

\begin{figure*}[!ht]
    \centering
    \includegraphics[width=\linewidth]{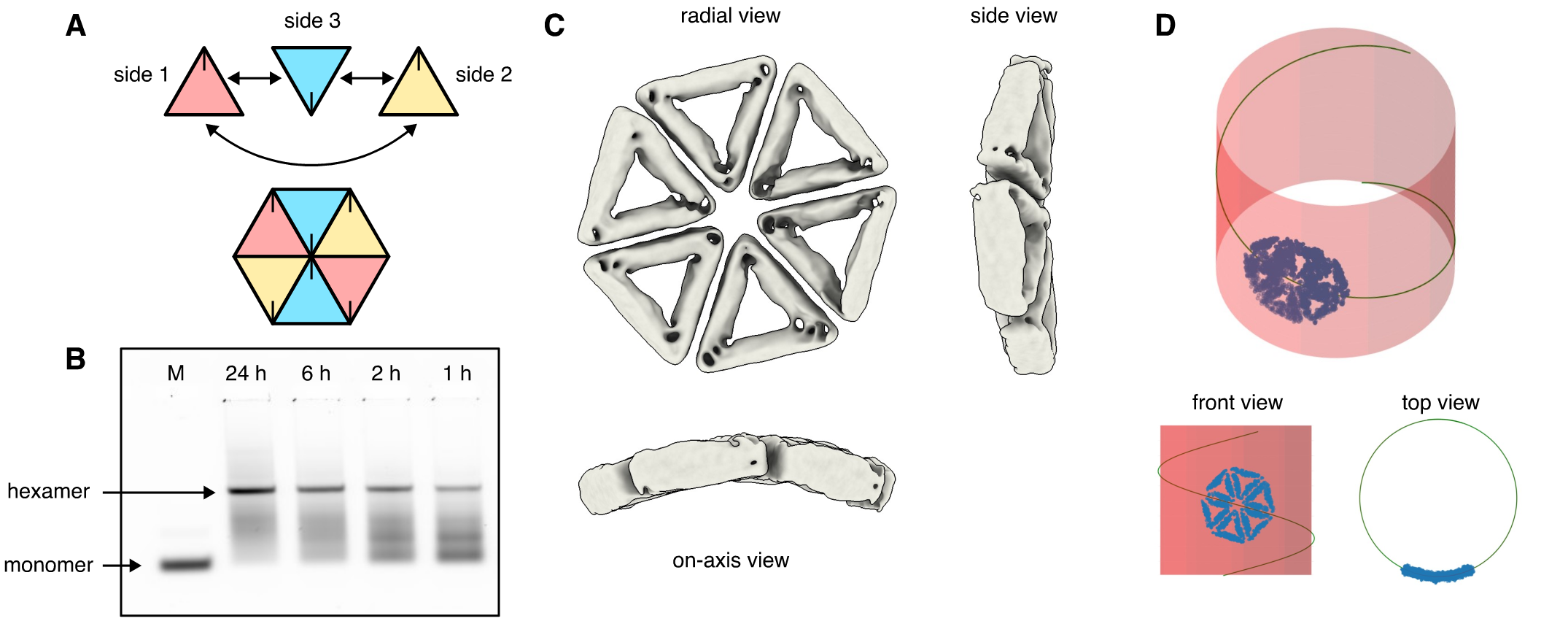}
    \caption{\textbf{Hexamer design and analysis}. (A) Design of hexamer using three particle species, the arrows between sides denote an attractive interaction. (B) Gel electrophoresis image for assembly of hexamers over time. A high yield of hexamers is seen after 24 hours. (C) Views of a hexamer reconstruction from three different views. The radial view shows that the side 3 seam has a slight pitch to it and the on-axis view shows the curvature of the hexamer. (D) A fitted cylinder to the hexamer. A subset of points from the hexamer (blue) are fitted to a cylinder (red surface). The side 3 seam is fit to a helix (green line). From and top views are also shown.}
    \label{Fig:sfig-hexamer-analysis}
\end{figure*}

In addition to the dimer reconstructions, we also perform cryo-EM imaging and reconstruction of a hexamer of triangular subunits. A hexamer is the smallest assembly of triangles where two bonds stabilize each particle, but due to the constraints of this assembly, the relative angles of neighboring triangles may vary from the angles of the individual dimers. To create a hexamer that preserves the orientations of triangles as they would be found in a tubule, we need to use three different monomer types (Fig.~\ref{Fig:sfig-hexamer-analysis}A). Once we have an assembled hexamer, we will measure the curvature of the assembly and compare it to the curvature of the tubules that are formed in experiment.

For preparing a sample for cryo-EM we need to make sure that stable hexamers have formed first. To check this, we mix the three triangle species in 1xFoB20 at 25~nM total DNA origami concentration and wait for 1, 2, 6, and 24 hours. After this, we prepare a 1\% agarose gel at 20 mM MgCl$_2$ look at the various time points of hexamer assembly (Fig.~\ref{Fig:sfig-hexamer-analysis}B). We find that after 24 hours there is a high yield of hexamers. To prepare a sample for plunging, we prepare a solution of 160 nM DNA origami (equal parts of each monomer type) in 1xFoB20 and wait for 24 hours.

After cryo-EM imaging and reconstruction, we measure the effective curvature of the hexamer assembly. From the average reconstruction, we can already see that the hexamer has qualities that are reminiscent of a patch of a tubule; there is a direction from which the projection appears as part of a circle, and an orthogonal view shows no curvature (Fig.~\ref{Fig:sfig-hexamer-analysis}C).

To measure the curvature of the hexamer we fit the electron density map to a cylindrical surface. To fit the electron density map we place a threshold on the electron density map and get the locations of all voxels that are above this threshold. We then take a subsample of these points to fit. A cylinder is defined by a center, an orientation axis, and a radius. For any fitting attempt, we take the points of the hexamer density map and cast them into cylindrical coordinates. Then we take the residual for the fit as the difference between the radial cylindrical coordinate and the fit radius. The best-fit parameters are found through least square minimization. To estimate uncertainties in the fit we use bootstrapping by performing this fitting procedure 100 times on different random subsamples of the hexamer points and take the standard deviation of the distributions for the fit parameters. The best-fit cylinder is shown overlaid with a single subsample of points from the hexamer density map in Fig.~\ref{Fig:sfig-hexamer-analysis}D.

To measure the pitch of the hexamer we fit a helix to the density of side 3 for the triangles. From the fit of our cylindrical surface, we have a best-fit center, orientation axis, and radius for the hexamer. This leaves only the determination of the orientation of the triangular lattice on the cylindrical surface. To determine this we use the eraser tool from ChimeraX and isolate the electron density from just side 3 of the triangles, the side that has the maximum angle with respect to the tubule axis. We then fit this density map to a helix on the surface of the cylinder that was fit to the hexamer. This best-fit helix has two parameters, its pitch and the phase. The best-fit helix is shown as the green curve in Fig.~\ref{Fig:sfig-hexamer-analysis}D. The best-fit parameters are a radius of curvature of 127 nm and a pitch of 223 nm. To use these to determine the nearest tubule type we also need to know the lattice spacing of the triangles. By measuring the distance between triangle tips in the hexamer we find a lattice spacing of 68.7 nm.

\clearpage

\begin{longtable}{p{0.06\linewidth} p{0.10\linewidth} p{0.10\linewidth} p{0.10\linewidth}p{0.10\linewidth} p{0.10\linewidth} p{0.10\linewidth} p{0.13\linewidth}}
\caption{\textbf{Side interactions subunits.} A list of the set of six interaction sequences that make up a side interaction of a monomer and an estimate of their summed binding free energy. The sequences are for self-complimentary side interactions, e.g. Position 1 binds to Position 6, Position 2 binds to Position 5, and Position 3 binds to Position 4.} \label{tab:sideinteractions} \\
\input{Tab-InteractionSeqs}
\end{longtable}

\begin{longtable}{p{0.06\linewidth} p{0.10\linewidth} p{0.10\linewidth} p{0.10\linewidth}p{0.10\linewidth} p{0.10\linewidth} p{0.10\linewidth} p{0.13\linewidth}}
\caption{\textbf{Side interactions for hexamer formation.} A list of the set of six interaction sequences that make up a side interaction of the three hexamer monomers shown in SFig.~\ref{Fig:sfig-hexamer-analysis}A and an estimate of their summed binding free energy. Sides of particles not listed have been passivated.} \label{tab:hexamer_interactions} \\
\input{Tab-HexamerInteractions}
\end{longtable}


\begin{figure*}[!h]
    \centering
    \includegraphics[width=\linewidth]{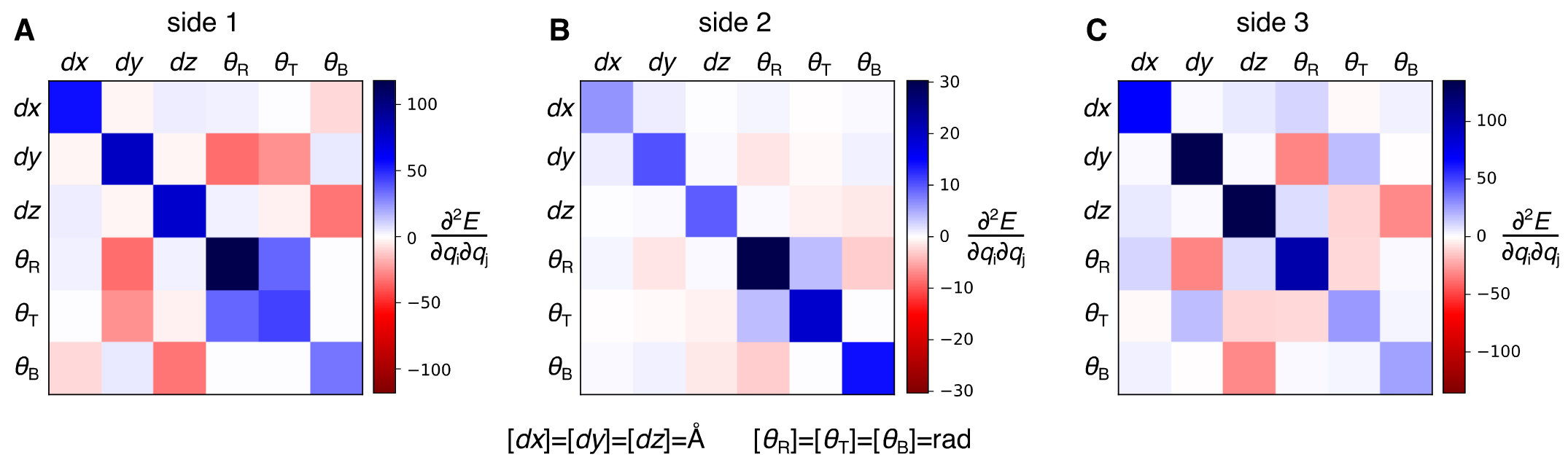}
    \caption{\textbf{Full dynamical matrices for all three sides.}. The panels are from the (A) side 1, (B) side 2, and (C) side 3 pairwise fluctuations measured in cryo-EM. Units for the displacements are in Angstrom and units for rotations are radians. All data has been scaled to account for the vitrification temperature.}
    \label{Fig:sfig-dynamical-matrices}
\end{figure*}


\begin{figure*}[!ht]
    \centering
    \includegraphics[width=\linewidth]{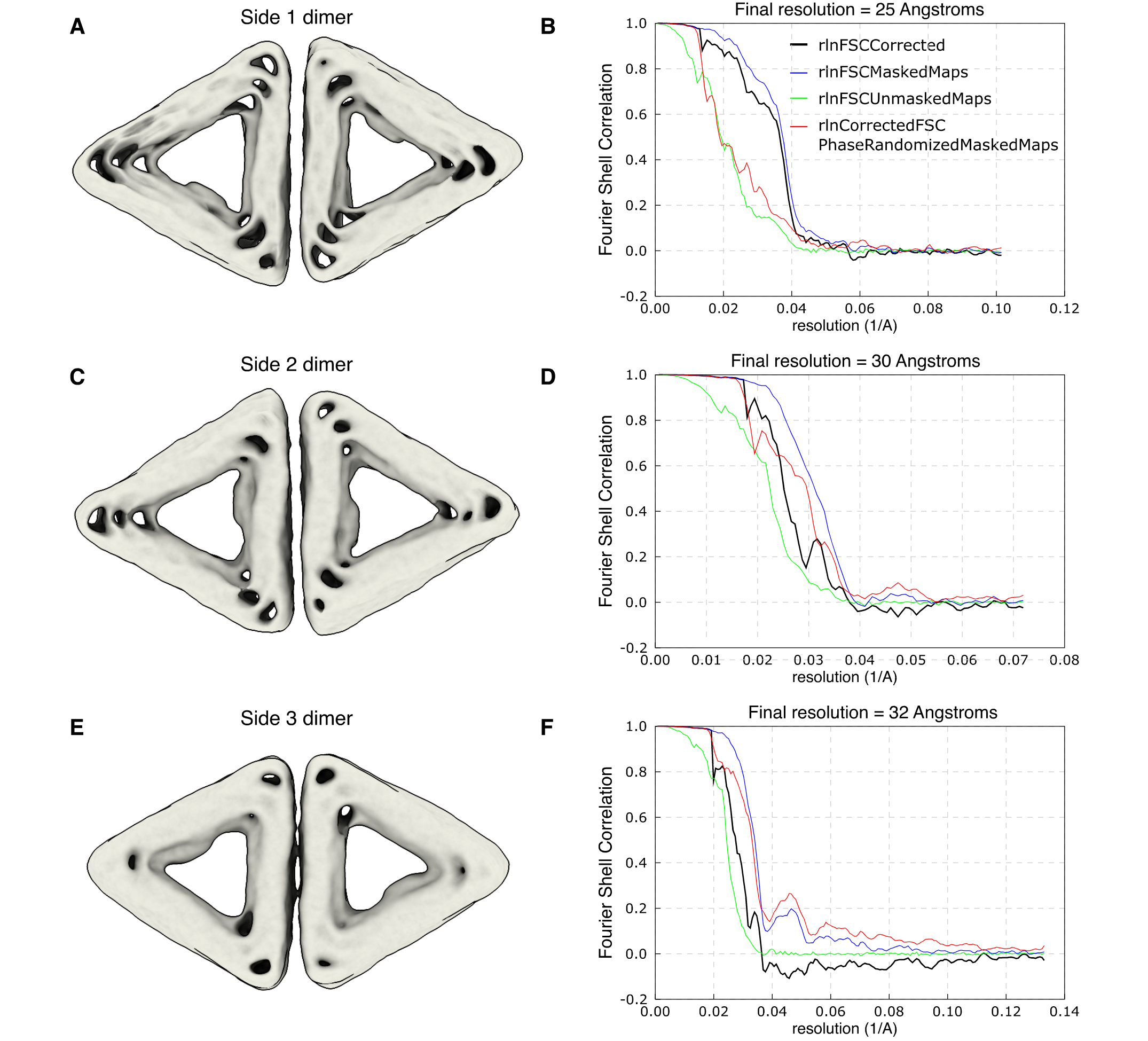}
    \caption{\textbf{Cryo-EM reconstructions of tubule-monomer dimers}. (A,C,E) Images of reconstructions for the dimers of triangles bound on sides 1, 2, and 3 respectively. (B,C,F) Plots of the FSC curves used to estimate the resolution of the dimer reconstructions.}
    \label{Fig:sfig-cryo-tubules-dimers}
\end{figure*}

\begin{figure*}[!ht]
    \centering
    \includegraphics[width=\linewidth]{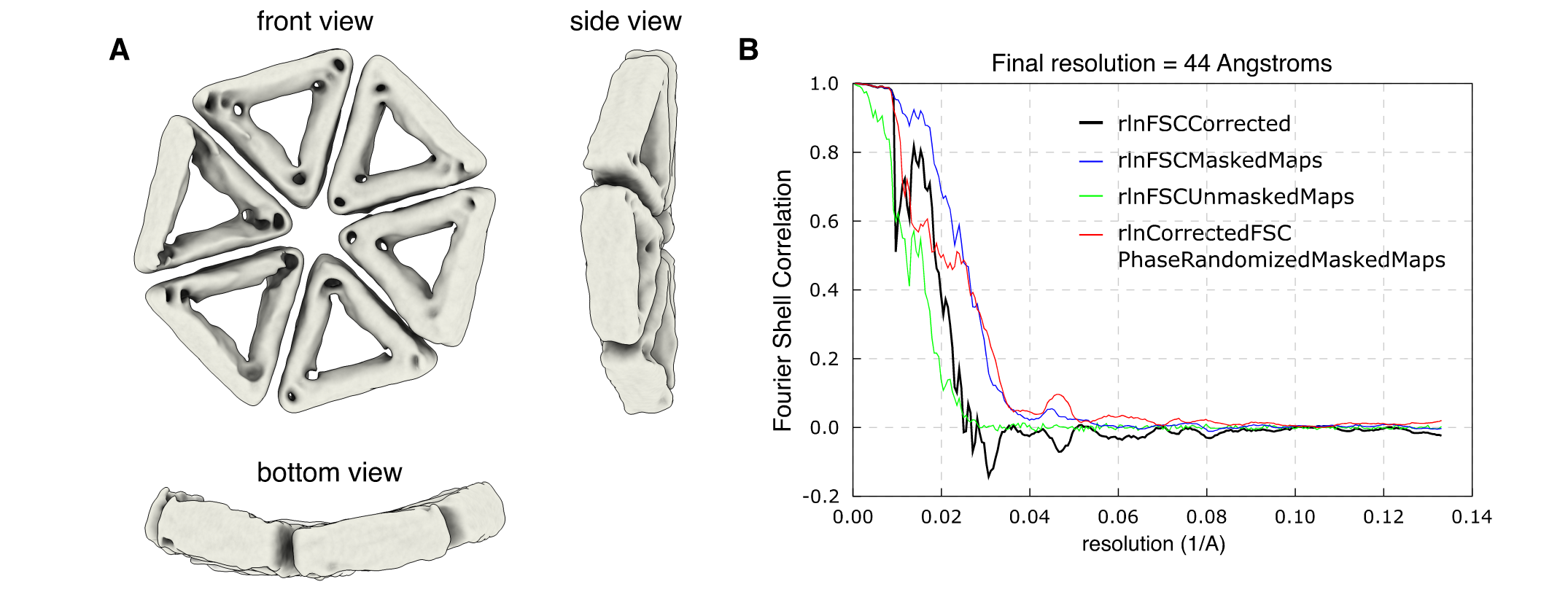}
    \caption{\textbf{Cryo-EM reconstruction of tubule-monomer hexamer}. (A) Different views show the average reconstruction of the hexamer. (B) Plot of the FSC curves used to estimate the resolution of the hexamer reconstruction.}
    \label{Fig:sfig-cryo-hexamer}
\end{figure*}

\begin{figure*}[!hbt]
    \centering
    \includegraphics[width=\linewidth]{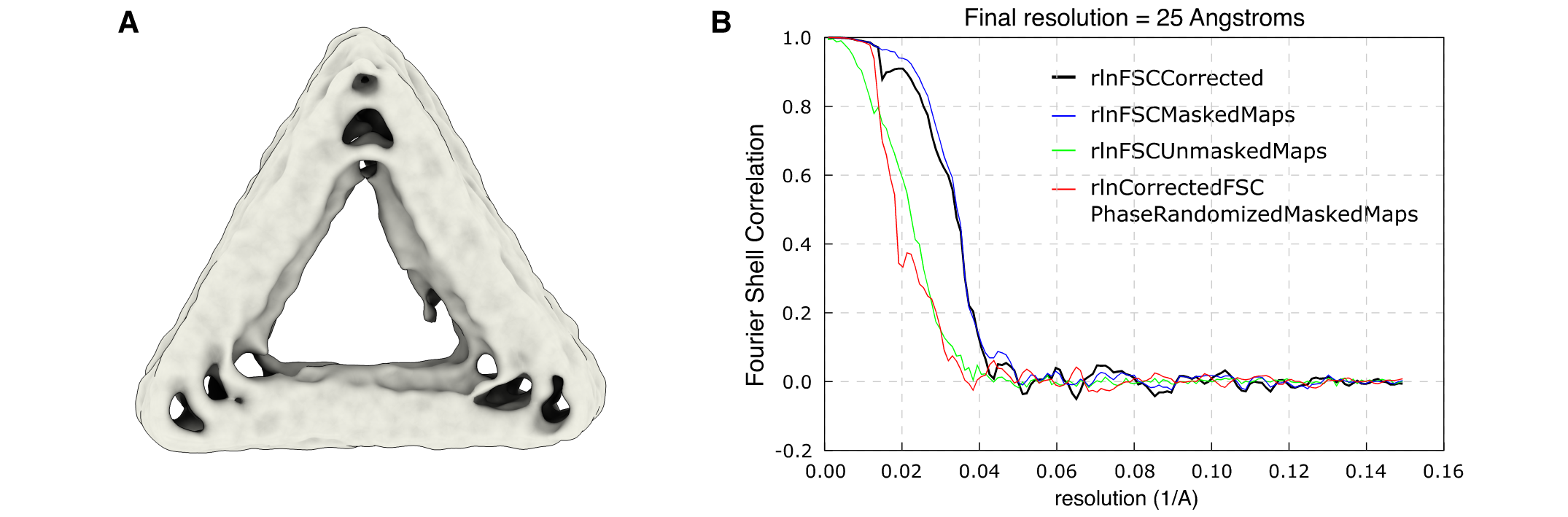}
    \caption{\textbf{Cryo-EM reconstruction of the twist-corrected monomer}. (A) Different views show the average reconstruction of the monomer. (B) Plot of the FSC curves used to estimate the resolution of the monomer reconstruction.}
    \label{Fig:sfig-cryo-mono-TC}
\end{figure*}

\clearpage

\begin{figure*}[!ht]
    \centering
    \includegraphics[width=\linewidth]{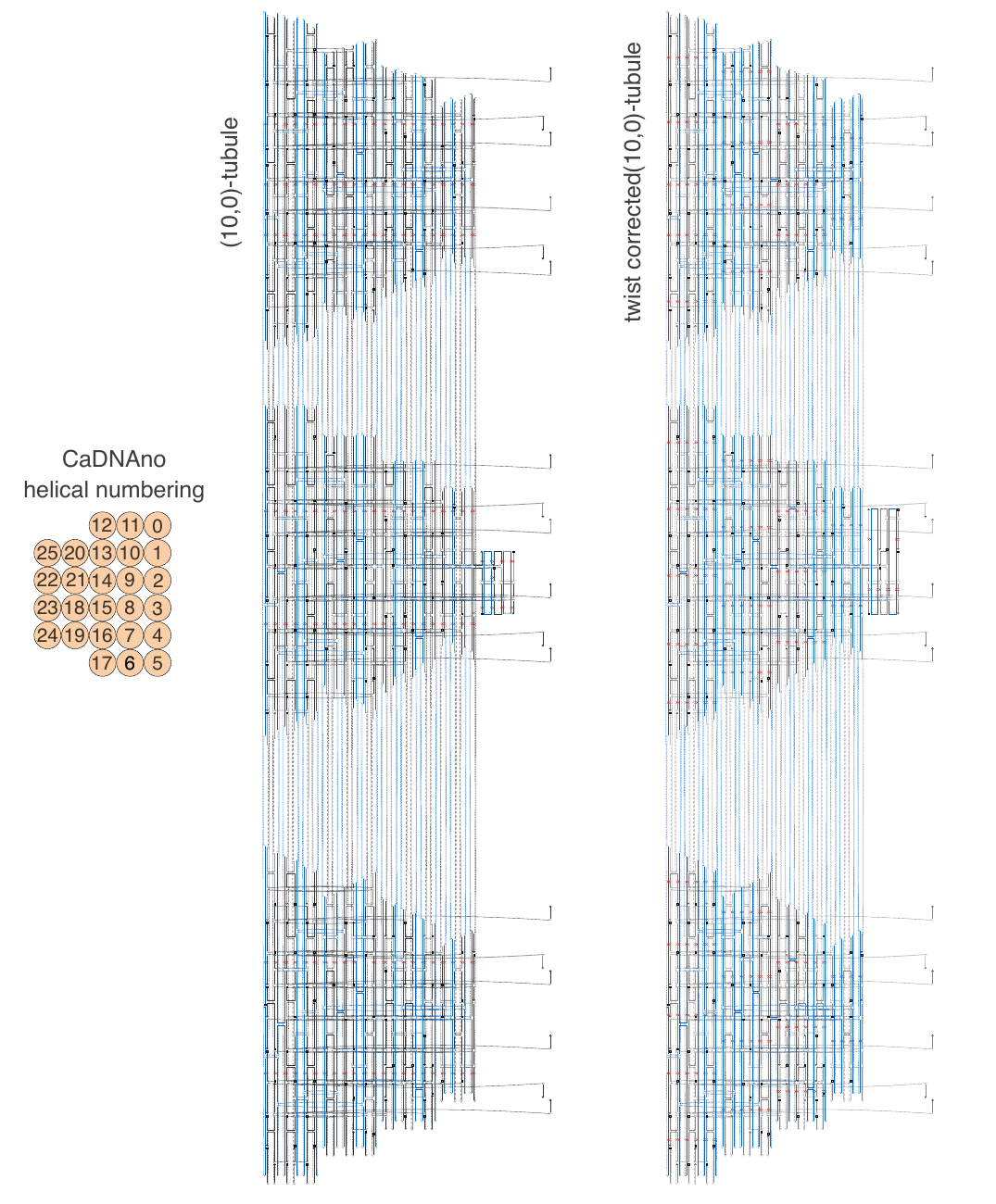}
    \caption{\textbf{caDNAno design of DNA origami monomer with helical numbering}. Red crosses are locations of skips in the design.}
    \label{Fig:sfig-cadnano}
\end{figure*}

\clearpage

\def\bibsection{\section*{Supplementary references}}

\bibliography{main.bib}

%% file: Tab-InteractionSeqs.tex
	&	Position 1	& 	Position 2	&	Position 3	&	Position 4	&	Position 5	&	Position 6	&	$\Delta$G [kcal/mol]	\\
\hline
\endfirsthead
 &	Position 1	& 	Position 2	&	Position 3	&	Position 4	&	Position 5	&	Position 6	&	$\Delta$G [kcal/mol]	\\
\hline
\endhead
Side 1	&	ACTAGC	& 	AGTTAC	&	TAGTCT	&	AGACTA	&	GTAACT	&	GCTAGT	&	-30.57	\\
Side 2	&	TTCAAT	& 	CCATTC	&	CTTGGT	&	ACCAAG	&	GAATGG	&	ATTGAA	&	-31.42	\\
Side 3	&	TTAACC	& 	TCGACA	&	TTGGAT	&	ATCCAA	&	TGTCGA	&	GGTTAA	&	-30.23	\\

%% file: Tab-HexamerInteractions.tex
	&	Position 1	& 	Position 2	&	Position 3	&	Position 4	&	Position 5	&	Position 6	&	$\Delta$G [kcal/mol]	\\
\hline
\endfirsthead
 &	Position 1	& 	Position 2	&	Position 3	&	Position 4	&	Position 5	&	Position 6	&	$\Delta$G [kcal/mol]	\\
\hline
\endhead

Blue side 1	&	CAATAG	& 	TGATTG	&	CTAGGA	&	CACATC	&	ACGAAG	&	ACCTGA	&	-31.38	\\
Blue side 2	&	ATGACA	& 	TACAGG	&	AACCTA	&	GAGACA	&	GACAGA	&	ACTAAC	&	-30.57	\\

Red side 1	&	TCAGGT	& 	CTTCGT	&	GATGTG	&	TCCTAG	&	CAATCA	&	CTATTG	&	-31.38	\\
Red side 3	&	GTACAT	& 	AGTCAG	&	CGATGG	&	CTTACT	&	AGTATC	&	GTATGT	&	-31.16	\\

Yellow side 2	&	GTTAGT	& 	TCTGTC	&	TGTCTC	&	TAGGTT	&	CCTGTA	&	TGTCAT	&	-30.57	\\
Yellow side 3	&	ACATAC	& 	GATACT	&	AGTAAG	&	CCATCG	&	CTGACT	&	ATGTAC	&	-31.16	\\